\documentclass[review]{elsarticle}

\usepackage{lineno,hyperref}
\modulolinenumbers[5]
\usepackage{lineno}
\journal{Journal of \LaTeX\ Templates}

\begin{document}

\begin{frontmatter}

\title{Study of encapsulated microbubble cluster based on association schemes perspective}

\author[T1]{S. Behnia\corref{mycorrespondingauthor}}
\cortext[mycorrespondingauthor]{Corresponding author}
\ead{s.behnia@sci.uut.ac.ir}

\author[B1]{M. Yahyavi\corref{mycorrespondingauthor2}}
\cortext[mycorrespondingauthor2]{Corresponding author}
\ead{m.yahyavi@bilkent.edu.tr}
\author[T1]{R. Habibpourbisafar}
\author[T2]{F. Mottaghi}
\address[T1]{Department of Physics, Urmia University of Technology, Orumieh, Iran.}
\address[B1]{Department of Physics, Bilkent University, 06800 Bilkent, Ankara, Turkey.}
\address[T2]{Department of Mechanical Engineering, Sahand University of Technology, Tabriz, Iran.}

\begin{abstract}
Ultrasound contrast agents have been recently utilized in therapeutical implementations for targeted delivery of pharmaceutical substances. Radial pulsations of a cluster of encapsulated microbubbles under the action of an ultrasound field are complex and highly nonlinear, particularly for drug and gene delivery applications with high acoustic pressure amplitudes. In this paper, based on Qin-Ferrara's model [S.P. Qin, K.W. Ferrara, J. Acoust. Soc. Am. 128 (2010) 1511], the complete synchronization and cluster formation in targeted microbubbles network are studied. Also, association schemes as a novel approach are suggested for finding a relationship between coupled microbubbles elements which are immersed in blood or surrounding soft tissue. { A significant advantage of this method is that the stability of the synchronized state (or symmetric eigenmode of mutual bubble oscillation) with respect to another state (another eigenmode) can now predict.} More interestingly, we find a significant relationship between an isolated and multiple microbubbles. The results show that the problem of studying the dynamics of encapsulated microbubble cluster at synchronization state is dependent on the dynamical characteristics of isolated cases, shell thickness, density. Also, the distance between microbubbles has an important role in their synchronous modes.
\end{abstract}

\begin{keyword}
Targeted drug delivery  \sep Encapsulated microbubbles \sep  Associated scheme \sep Bose-Mesner algebra \sep Globally coupled map \sep Synchronization
\sep Lyapunov exponent
\end{keyword}

\end{frontmatter}

\linenumbers

\section{Introduction}
Ultrasound contrast agents (UCAs) are spherical microbubbles which were coated by a shell such as albumin or polymer and filled with a high-molecular-weight gas. Recently UCAs with small diameters (1-5$\mu$m) have been utilized in clinical application and their employment in the biomedical field is promoting to the therapeutic applications such as targeted drug and gene delivery~\cite{sirsouy,casfrg}. Many sophisticated theoretical models have been studied for describing an isolated UCAs motion. Fundamental perception of UCAs dynamics and precisely predicting their behavior will promote their diagnostic and therapeutic capabilities; indeed a quantitative understanding of UCAs dynamics is a necessary step to attain a better hardware design and successful clinical applications. Many sophisticated theoretical treatments for describing the coated microbubble response in an ultrasound field have been performed whereas most of the presented models are on the foundation of the Rayleigh-Plesset (R-P) equation form. De Jong and co-workers~\cite{delad} introduced the first theoretical model that considers the encapsulation as a viscoelastic solid shell, as well as a damping coefficient term, is added to the R-P equation. Church~\cite{chur} presented a more accurate model by considering the shell thickness to describe the effects of the shell on UCA behavior. Morgan (see also Allen model for modified Herring equation)~\cite{morga,allen} and Zheng~\cite{zheng} offered their models for thin and thick encapsulated microbubbles, respectively. Another rigorous model which treats the outer medium as a slightly compressible viscoelastic liquid is due to Khismatullin and Nadim~\cite{khismatullin}. Chatterjee and Sarkar~\cite{sarkar2} attempted to take account of the interfacial tension at the microbubble interface with infinite small shell thickness. Sarkar~\cite{sarkar} improved this model to contain the surface elasticity by using a viscoelastic model. Stride and Saffari~\cite{stride} demonstrated that the presence of blood cells and the adhesion of them to the shell have a negligible effect. Marmottant~\cite{marmo} exhibited a simple model for the dynamics of phospholipid-shelled microbubbles while taking account of a buckling surface radius, shell compressibility, and a break-up shell tension. Doinikov and Dayton~\cite{doinik} refined the Church model and also considered the translational motion of the UCA. Shengping Qin and Katherine W. Ferrara~\cite{qinnm} have presented a model to explain the radial oscillations of UCAs by considering the effects of liquid compressibility, the surrounding tissue, and the shell. The effects of radially asymmetric shell elasticity where shell viscosity is radially dependent on the modified R-P equation have been investigated by Helfield et al.~\cite{fostertt}.

 In fact, clinical applications of UCAs often involve multiple agents in clusters~\cite{boukazs,kruser,postema}. A cluster of microbubbles is defined as a group of two or more microbubbles, which interacted together. {\bf The collective behavior of microbubbles is different from the isolated case.
The simplest difference is due to the acoustic coupling between bubbles \cite{Crum,Doinikov02,Feuillade01,N19,mett,Doinikov3}  
that can occur irrespective of whether the bubbles are translating, and in an incompressible
liquid. This is the focus of the present paper.}
 
 The first model was theoretically studied by Takahira~\cite{takahira} which is demonstrated ensemble dynamics of free gas bubbles. Also, by using Lagrangian equations motion and an extension of the R-P equation, the effect of coupled oscillation on two interacting encapsulated microbubbles with different initial radius have been investigated~\cite{alllem}. The behavior of thin encapsulated microbubble in a cluster has been studied which takes into account the elastic properties of the UCAs shell~\cite{macdonbnh}.

On the other hand, UCAs (single or cluster) undergo complex dynamic behaviors while they are exposed to an ultrasound field, which depending on the applied acoustic amplitudes, they will respond linear~\cite{alllemasd,hoffgt} or nonlinear pulsations~\cite{alllem,behnia2,behnia3,behnia4,behnia5,dzah}. The nonlinear nature of the equations needs specialized tools for analyzing due to the fact that linear and analytical solutions are inadequate. Obviously, in order to be close to the real case, analysis of the microbubble cluster are required.  {Since the linear and other suggested analytical solutions are not enough for representation the nonlinear behaviors of above models then bubble dynamics for analyzing needs to specialized tools. Hence, the chaotic motion of isolated and interacted bubbles observed theoretically by chaos theory tools such as bifurcation and Lyapunov diagrams. The mentioned methods have been used to study the bubble cluster's behavior with only a finite number of elements (3 or 4 bubbles). Since the Qin-Ferrara's model for UCAs is usually not studied in terms of $N$ interacting albumin and polymer agents, hence the question arises, the coupled map lattices (CML) gives information of the stability region of which physical quantity? Can one investigate the dynamic behaviors of $N$ interacting agents cluster from the CML approach at complete synchronization state?} To answer to this question, the following study is provided. At the first step, the globally coupled map (GCM) model for encapsulated microbubbles is introduced. { It was emphasized, that an extension of CML, in which each element is connected with all other elements is called GCM.} Then, by generating the corresponding graph, the basis for utilization algebraic structure of graph theory (association schemes theory) is prepared. In the next step, we have tried to recognize the synchronization condition in the proposed mathematical model. It is shown that the stability analysis at the synchronized state reduces to study the dynamics of the isolated microbubble. Dynamical behavior of single microbubble by considering its chaotic nature have been studied by employing chaos theory tools such as the bifurcation and Lyapunov exponent diagrams~\cite{sojahro,carrklo}.

{To answer the above questions, we will employ an association scheme in order to calculate the $N$ microbubble cluster stability at complete synchronization state. The rest of this paper is organized as follows: In the next section (Sec. \ref{Sec2}), we introduce the concept of CML which is one significant method to investigate the emergent phenomena such as synchronization, cooperation, and more, which may happen in interacting physical systems. A theoretical model for $N$-microbubbles interaction is introduced in Sec. \ref{Sec3},  and also we employed an association scheme algebra in order to calculate the stability condition of a composed cluster of the albumin and polymer agents in the cluster at complete synchronization state.  In Sec. \ref{Sec4} our results and analyses are presented. In Sec. \ref{Sec5} we conclude our work. Three appendices are also provided, which contain all association scheme calculations and proofs.
}

\section{Stability analysis based on Lyapunov exponent of $N$-coupled dynamical systems}\label{Sec2}
{One of the active fields in coupled nonlinear dynamical systems is
synchronization. Synchronization of chaotic coupled dynamical systems with different initial conditions means that all elements trajectories remain with together in any time step. One of the powerful mathematical technique which has been used in analyzing the synchronization state is the relation between the array of coupled system and graph theory (or CML approach).} The GCM as an extension of CML consists of an array of dynamical elements (nodes) where each element interacts with all the  other elements and their values are continuous or discrete in space and time~\cite{kan1}. So consider a network of $N$ nodes and $N$ couplings between them, which each node of the network can be characterized as a dynamical variable $x_{i}$. Then, GCM model can be written as:
\begin{equation}\label{global}
\dot{x_{i}}(t)= (1-\epsilon)f(x_{i}(t))+\frac{\epsilon}{N}\sum_{j=1,j \neq i}^N f(x_{j}(t))
,\quad
 i,j=1,\cdot\cdot\cdot,N
\end{equation}
where $x_{i}(t)$ is the state variable of node $i$ (vertex $i$). {The first term at the right-hand-side of above equation represents the unit's dynamics given by a properly-selected nonlinear mapping function $f(x)$ while the second term is the coupling of the interaction through
the coupled parameter $\epsilon $ $(0 \leq  \epsilon  \leq 1)$. When $\epsilon $  is small, it is weak coupling and the strength of
coupling increases with $\epsilon $, i.e. $\epsilon=1$ is the largest coupling of dynamics and  the dynamics are simple when a lattice formed by
independent, uncoupled ($\epsilon=0$) elements.} Functions $f(x_{i})$ and $f(x_{j})$ describe the single dynamical system and interaction between the individual nodes, respectively. {Association scheme theory has a capability to explain the relation between the pairs of the elements of the bubbles. However, in order to perform the association schemes perspective on interacting bubbles cluster, we consider a coupling matrix  to create a connection between the GCM model and association schemes theory, }
\begin{equation}\label{mesnerrr}
\mathcal{A}=\sum_{\alpha=0}^d \frac{\epsilon(\alpha)}{\kappa(\alpha)} A(\alpha)
\end{equation}
where $A(\alpha)$ is the adjacency matrix of $\mathcal{B}(\alpha)$ (\ref{AppendixA}) { and $\kappa$ is the closeness of another graph member (bubble$-j$)  to graph member (bubble$-i$)}. {In fact, the coupling matrix could generate well-known Markovian matrices, which each element based on $\epsilon({\alpha})$ in $\mathcal{A}$ is transition probability from vertex $i$ to $j$ in the complete graph. Also, $\epsilon(\alpha)$ is directly connected to numbers of  physical parameters of bubbles (please see Eq. \ref{froha}). We have study the behavior of bubble-bubble interaction at complete synchronization state.  As can see, the Eq. (\ref{mesnerrr}) with generating Markovian matrix have following stationary transition probabilities}
\begin{equation}
\mathcal{A}=\left(
 \begin{array}{cccc}
 1-\epsilon &  \epsilon & \cdots & \sum_{\alpha=1}^{d}\frac{\epsilon(\alpha)}{\kappa(\alpha)} \\
 \epsilon &  1-\epsilon & \frac{\epsilon}{2} & \cdots \\
 \vdots  & \frac{\epsilon}{2} & 1-\epsilon & \cdots  \\
 \sum_{\alpha=1}^{d}\frac{\epsilon(\alpha)}{\kappa(\alpha)} & \vdots & \cdots & \ddots
 \end{array}
 \right)
\end{equation}
{In this simple and strong form of synchronization, with existence an invariant sub-manifold, we have $x_{1}=\cdot \cdot \cdot =x_{N}$, which all state variables of coupled dynamical systems have identical behaviors and are present in synchronization manifolds.} Therefore, by considering that $\epsilon(0)=1-\epsilon$ and $\epsilon(1)=\epsilon$, we have
\begin{equation}\label{synch}
\dot{x_{i}}(t)=\sum_{j=1}^N \mathcal{A} f(x_{j}(t)),\quad   j= 1, 2, ..., N
\end{equation}
On the other hand, the GCM model with global coupling strength corresponds to the complete graph~\cite{gup}. Indeed, a complete graph $K_N$ with $N$ vertices and $\frac{N(N-1)}{2}$ edges is a bidirectional graph where each pair of distinct vertices is connected by an unique edge (see Fig. 1). So $A(\alpha)$ are adjacency matrices of the complete graph which can be cover the topology of the GCM model.
Moreover, coupling parameters $\epsilon(\alpha)$ with respect to Eq. (\ref{global}) satisfy
\begin{equation}\label{mosavi}
\sum_{\alpha=0}^{d}\epsilon(\alpha)=1
\end{equation}
obviously, the sum of columns and rows of $\mathcal{A}$ is one. In fact, coupling matrix could generate well-known Markovian matrices~\cite{senet}, which each element based on $\epsilon(\alpha)$ in $\mathcal{A}$ is transition probability from vertex $i$ to $j$ in the complete graph. { We have studied the behavior of the bubble-bubble
interaction at complete synchronization state. } Synchronization of two or more chaotic dynamical systems in a coupled network, which starts with different initial conditions, means that their trajectories remain in step with each other during the temporal evolution ~\cite{wu}. When one deals with identical systems, the specific synchronized motion happens which called complete synchronization (CS). In a geometric representation, CS corresponds to the collapse of the full network's trajectory in the phase space onto an identity hyperplane which called synchronization manifold, i.e. $x_{1}(t)=\cdot \cdot \cdot=x_{N}(t)$. {It is important to notice that a necessary condition for stability the synchronization manifold is that the set of Lyapunov exponents that corresponds to phase space directions transverse to the {identity  hyperplane} be entirely made of negative values.} In order to analyze the stability at synchronized state by perturbing
$$
\delta \dot{x_{i}}(t)= \sum_{j=1}^N \left(\frac{\partial \dot{x_{i}}(t)}{\partial x_{j}(t)}\right)_{x_{1}(t)=\cdot \cdot \cdot=x_{N}(t)}\delta x_{i}(t)
$$
and by taking derivative of $\dot{x_{i}}(t)$ with respect to $x_{j}(t)$ in Eq. (\ref{synch}), we obtain
$$
\delta \dot{x_{i}}(t)= \mathcal{A} \frac{\partial f(x_{i}(t))}{\partial x_{i}(t)} \delta x_{i}(t)
$$
By iterating
\begin{equation}\label{itera}
\delta \dot{x_{i}}(t)=\left(\prod _{n=0}^{t-1} \mathcal{A} f'(x_{i}(n))\right)\delta x_{i}(0)=(\mathcal{A})^n
\left(\prod _{n=0}^{t-1}f'(x_{i}(n))\right) \delta x_{i}(0)
\end{equation}
with substituting Eq. (\ref{eigen1}) in Eq. (\ref{mesnerrr}), we have
$$\mathcal{A}=\sum_{\beta=0}^d \left(\sum_{\alpha=0}^d \frac{\epsilon(\alpha)}{\kappa(\alpha)} P(\alpha,\beta)\right) S(\beta)$$
Then
$$
(\mathcal{A})^n=\sum_{\beta=0}^d \left(\sum_{\alpha=0}^d \frac{\epsilon(\alpha)}{\kappa(\alpha)}P(\alpha,\beta)\right)^{n} S(\beta)
$$
Now, Eq. (\ref{itera}) can be written as:
$$
\delta \dot{x_{i}}(t)=\sum_{\beta=0}^d \left(\sum_{\alpha=0}^d \frac{\epsilon(\alpha)}{\kappa(\alpha)}P(\alpha,\beta)\right)^n
\left(\prod _{n=0}^{t-1} f'(x_{i}(n))\right) S(\beta) \delta x_{i}(0)
$$
By considering $\beta=0$, we have
\begin{equation}\label{lyaaa}
\sum_{\alpha=0}^d \frac{\epsilon(\alpha)}{\kappa(\alpha)}P(\alpha,0)=1
\end{equation}
that leads us to write
$$\delta \dot{x_{i}}(t)=\left(\prod _{n=0}^{t-1} f'(x_{i}(n))\right)S(0) \delta x_{i}(0)+ \sum_{\beta=1}^d \left(\prod _{n=0}^{t-1} f'(x_{i}(n))\right)
\left(\frac{\epsilon(\alpha)}{\kappa(\alpha)}P(\alpha,\beta)\right)^{n} S(\beta) \delta x_{i}(0)
$$
where $S(0)$ $\delta x_{i}(0)$ represents the synchronous mode and $S(\beta)$ $\delta x_{i}(0)$ is dependent on transversely state.  {In the language of a dynamical system, presentation of $S(0)$ means a complete synchronization of the dynamical systems}

\begin{equation}\label{S0}
 S(0)=\frac{1}{N}J_{N}=\frac{1}{N}\left(
                                     \begin{array}{ccccc}
                                       1 & 1 & \cdot & \cdot & 1 \\
                                     \end{array}
                                   \right)
\left(
                                     \begin{array}{c}
                                       1 \\
                                       1 \\
                                       \cdot \\
                                       \cdot \\
                                       1 \\
                                     \end{array}
                                   \right)
\end{equation}
{the other element of $ S(\beta)$ project the system to transverse mode.}   So the Lyapunov exponent of $N$-coupled dynamical systems are defined as
$$
\Lambda(\beta)=\lim_{t\longrightarrow \infty} \frac{1}{t} \ln \frac{\|\delta \dot{x_{i}}(t) \|}{\|\delta x_{i}(0)\|}
= \lambda_{f(x_{i}(t))}+ \ln \left|\sum_{\alpha=0}^d \frac{\epsilon(\alpha)}{\kappa(\alpha)}P(\alpha,\beta)\right|
$$
with respect to Eq. (\ref{lyaaa}), $\lambda_{f(x_{i}(t))}$ shows the Lyapunov exponent for single dynamical system. For the stability of transverse mode it is necessary to have $\Lambda(\beta)<0 $, so extremum bound on synchronous state is given by
\begin{equation}\label{lyappa}
1-e^{-\lambda_{f(x_{i}(t))}} \leq \sum_{\alpha=1}^{d} \frac{\epsilon(\alpha)\left[\kappa(\alpha)-P(\alpha,\beta)\right]}{\kappa(\alpha)}
\leq 1+e^{-\lambda_{f(x_{i}(t))}}, \quad   \beta=1,2,\cdot \cdot \cdot, d
\end{equation}
The stable domain at CS state is restricted between the coupling $\epsilon(\alpha)$ and association schemes parameters.

As a consequence of this process, generally, the appearance of CS is accompanied by an effective abrupt decrease of the dimensionality
(complexity) of the overall motion. {Once again, it has to be remarked that the stability condition of Eq. (\ref{lyappa}) define a hyper parallelogram in hyperspace to be formed by coupling parameter, $\epsilon(\alpha)$. The created hyperplane by synchronization manifold corresponds to the group of bubbles which oscillate with the same phase. As a consequence of this process and approach, generally, the appearance of CS is accompanied with an effective abrupt decrease of the dimensionality of a cluster of the bubble (see Eq. (\ref{S0}) and (\ref{lyappa})). Above inequality shows that properties of the single system (i.e. isolated bubble) can be affected
the synchronized state. Clearly, stability influenced with coupling strength.} The number of vertices and associate classes have an important role in the inequality condition. By increasing the number of microbubble (vertices), the stability region is reduced and their trajectories are no converge to the same value where coupled dynamical systems are in the synchronization state.
\section{Theoretical model for targeted microbubble cluster}\label{Sec3}
The $N$ interacting targeted microbubbles cluster model, which is immersed in blood or tissue, on the basis of the Qin and Ferrara equation is given by~\cite{qinnm,takahira,mett}
$$\left[\rho_{L}\left(1-\frac{\dot{R}_{2i}}{c}\right)+\rho_{E}\left(1+\frac{\dot{R}_{2i}}{c}\right) \left(\frac{R_{2i}}{R_{1i}}-1\right) \right] R_{2i}\ddot{R}_{2i} +\left\{\frac{3}{2}\rho_{L} \left(1-\frac{1}{3}\frac{\dot{R}_{2i}}{c}\right)+\right.$$
 $$
\left. \rho_{E} \left(1+\frac{\dot{R}_{2i}}{c}\right) \left[-\frac{3}{2}+2 \left(\frac{R_{2i}}{R_{1i}}\right)-\frac{1}{2}\left(\frac{R_{2i}}{R_{1i}}\right)^4\right] \right\}\dot{R}_{2i}^2=\left(1+\frac{\dot{R}_{2i}}{c}\right)\left\{ p_{gi}(t)-\frac{2\sigma_1}{R_{1i}}-\right. $$
$$ \left. \frac{2\sigma_2}{R_{2i}}-\frac{4}{3}G_{E}\left[1-\left(\frac{R_{2i0}}{R_{2i}}\right)^3\right]\frac{W_{E}}{R_{2i}^3-W_{E}}-
4\mu_{E}\frac{W_{E}}{R_{2i}^3-W_{E}}\frac{\dot{R}_{2i}}{R_{2i}}-\frac{4}{3}G_L \left[1-\left(\frac{R_{2i0}}{R_{2i}}\right)^3\right]-\right.$$
\begin{equation}\label{Microbubble}
\left. 4\mu_L\frac{\dot{R}_{2i}}{R_{2i}}-p_0-
P_{ac}\sin (2\pi \nu t)-\sum_{j=1,j\neq i}^{N}\frac{\rho_{L}}{D_{ij}}\frac{d}{dt}\left(R_{2j}^{2}\dot{R}_{2j}\right)\right\}-3\Gamma\frac{\dot{R}_{2i}}{c}
\left(\frac{R_{2i}}{R_{1i0}}\right)^3\frac{p_{gi}(t)}{\left(\frac{R_{1i}}{R_{1i0}}\right)^3-\frac{q}{w_{m}}}
\end{equation}
where
\begin{equation}
R_{1i}=\left(R_{2i}^{3}-W_{E}\right)^{\frac{1}{3}}, \quad W_{E}=R_{2i0}^{3}-R_{1i0}^{3}, \quad p_{gi}(t)=\left(p_{0}+\frac{2\sigma_{1}}{R_{1i0}}+\frac{2\sigma_{2}}{R_{2i0}}\right)\left[\frac{1}{\left(\frac{R_{1}}{R_{1i0}}\right)^{3}-\frac{q}{w_{m}}}\right]^{\Gamma}
\end{equation}
Changing subscripts $i$ and $j$ yields $j$-th UCAs.
In the above equation, $R_{1}$ ($R_{2}$) is the inner (outer) radius of the agents, $\dot{R}_{1}$ ($\dot{R}_{2}$) is the inner (outer) wall velocity of the agents and $\ddot{R}_{2}$ is the outer wall acceleration of the agents.   {The subscript $``0"$ represents the equilibrium state.} The $p_{ac}$ is the ultrasound pressure at infinity and $\nu$ is the ultrasound center frequency, $D_{ij}$ is the distances between microbubbles and $N$ denotes the number of interacting microbubbles, which confined in a cluster. {In above equation $\sigma_1$ is the inner surface tension, $\sigma_2$ is the outer surface tension, $G_s$ is the shear modulus,  $c$ is the speed of sound in the liquid,  $q$ is the van der Waals constant, $w_m$   is the universal molar volume, $\rho$ is the density of the liquid, $\mu$ is the viscosity of the liquid.} The subscripts $``E", ``L"$ refer to the shell and surrounding fluid properties, respectively. The parameter values for albumin and polymer agents are summarized in Table 1. The shear modulus of the medium is $G_{L}$=0, 0.5, 1.5 MPa. The Qin-Ferrara cluster model represents GCM model in which each outer radius $R_{2i}(t)$ replace with vertices of the complete graph.

 {In order to calculate the $N$ interacting targeted microbubble cluster of Eq. (\ref{Microbubble}), $N+1$ equations should be simultaneously solved. However, numerous attempts have been made in the literature~\cite{Yasui08,Yasui09},  to reduce the number of the equations to be solved for spatially homogeneous distribution of bubbles. { It has been shown in ~\cite{Yasui08,Yasui09} that when the pressure amplitude of ultrasound and ambient bubble radius is the same for all the bubbles, the pulsation of all the homogeneously distributed bubbles will be in the same manner.}  It is worth mentioning that for some value of coupling strength, the effect of the bubble-bubble interaction is nearly negligible, \cite{Yasui10}. It was emphasized, that due to the acoustic coupling of the bubbles, a bubble chain itself will cause an anisotropy in the sound field. Subsequently, Feuillade,  using a coupled-oscillator or self-consistent model, conducted a mathematical analysis of the radiation damping due to the interaction of the bubble in  two- and three-bubbles systems over a separation range. In the same article, Feuillade \cite{Feuillade01} showed that for closely spaced bubbles oscillating in phase, the interaction of the bubble effect in a compressible liquid tends to increase damping, and for anti-phase oscillations tend to reduce damping. As further bubbles are added to the system, this method is analytically problematic. Actually, for  many bubbles, it can be shown, by setting up larger sets of coupled differential equations.  }

 \textbf{Based on previous studies, when bubbles in the cluster interact with each other, their ensemble dynamics are remarkably different from an isolated bubble, for example, bubbles may be repelled or attract one another (secondary Bjerknes force) \cite{Crum,Doinikov02,Feuillade01}.      According to the linear theory \cite{N19,mett,Doinikov3}, the secondary Bjerknes force between
the two bubbles is repulsive, when the driving frequency is between the linear resonance frequencies of the two bubbles. Otherwise, the secondary Bjerknes
force is attractive. } {In this paper, the only radial motion of the bubbles was taken into account. However, bubble translation occurs due to Bjerknes forces associated with the presence of nearby surfaces, the pulsation of neighboring bubbles, and the gradients in the acoustic field \cite{BB1}.} {The assumption used in this paper is that the distance $D_{ij}$ between the bubble $i$ and bubble $j$ is large enough to ensure that the bubbles throughout their motion remain spherical and the surrounding liquid is incompressible and {we neglected the translational motion of bubbles due to the secondary Bjerknes
force (an interaction force proportional to $D_{ij}$), that is, assuming $\frac{dD_{ij}}{dt} = 0$.} Following the Ref. \cite{mett,behnia2}, there is a time-delay $\tau=\frac{D_{ij}}{c}$ that occurs between an oscillation of bubble $i$ and the action of the incremental pressure/density on the bubble $j$ at a distance $D_{ij}$. The source of the time-delay comes
from the time it takes for the signal to travel from one
bubble to the neighbor spherical bubble through the liquid medium which
surrounds them at a distance $D_{ij}$.  Here, we set the
distance $D_{ij}$ between the $i$th and $j$th bubble to be much larger than
$R_{2i0}$ and by comparing the maximum value of $\tau$ and $T = \frac{1}{\nu}$,
one can achieve $\tau < T$, which means that the time-delay has an insignificant
effect on the result of this paper. Therefore the compressibility of the liquid
is negligible and the approximation of an incompressible
fluid has been used.       }

\subsection{stability region of UCAs cluster}
The capability of introduced algebra by considering the targeted microbubble cluster with the different shell (\ref{AppendixB})
and three elements (see Fig. 2) is examined and explained in a wide range of the parameter domain. So let we consider two different types of three interacting UCAs clusters (albumin and polymer shell) which arranged in vertices of the equilateral triangle, where ($D_{12}=D_{13}=D_{23}=D$) for the albumin shell and ($D'_{12}=D'_{13}=D'_{23}=D'$) for Polymer agents.

Each cluster has two associated class $\{\mathcal{B}_{1}(0),\mathcal{B}_{1}(1)\}$ and $\{\mathcal{B}_{2}(0),\mathcal{B}_{2}(1)\}$ with scheme class $0\leq\alpha_{1}\leq1$ and $0\leq\alpha_{2}\leq1$, where subscript $"1"$ refers to the albumin and $"2"$ denotes the polymer shell. The corresponding valencies of two association schemes with respect to Eq. (\ref{sareza}) and $N_{1(2)}=|V_{1(2)}|=3$ are
\begin{equation}\label{madelk}
\kappa_{1(2)}(0)=1, \quad \kappa_{1(2)}(1)=2
\end{equation}
The coupling matrices of two clusters with different properties of shells and distances in the framework of Eq. (\ref{mesnerrr}) with respect to Eq. (\ref{Microbubble}) are
\begin{equation}\label{markovooo}
\mathcal{A}_{1}=a A_{1}(0)+ \frac{b}{2} A_{1}(1), \quad \mathcal{A}_{2}=a' A_{2}(0)+ \frac{b'}{2} A_{2}(1)
\end{equation}
where
$$
b=\frac{\rho_{L}D}{ \left(\frac{\rho_{E}^{2}D^{2}}{W_{E}^{\frac{2}{3}}}-2\rho_{L}^{2}-\frac{\rho_{L}\rho_{E}D}{W_{E}^{\frac{1}{3}}}\right)}, \quad  a=1-b
$$
\begin{equation}
b'=\frac{\rho_{L}D'}{ \left(\frac{\rho_{E}^{'2}D^{'2}}{W_{E}^{'\frac{2}{3}}}-2\rho_{L}^{2}-\frac{\rho_{L}\rho_{E}'D'}{W_{E}^{'\frac{1}{3}}}\right)}, \quad
a'=1-b'
\end{equation}
and
\begin{equation}
A_{1(2)}(0)= \left(
                   \begin{array}{cccc}
                     1 & 0 & 0  \\
                     0 & 1 & 0  \\
                     0 & 0 & 1  \\
                   \end{array}
                 \right)=I_{3} , \quad
A_{1(2)}(1)=\left(
                     \begin{array}{cccc}
                       0 & 1 &1  \\
                       1 & 0 & 1  \\
                       1 & 1 & 0  \\
                     \end{array}
                   \right)
\end{equation}
where rows and columns of these matrices are corresponding to the vertices of the complete graph.
The $\mathcal{A}_{1}$ and $\mathcal{A}_{2}$ could generate Markovian matrices. For instance, entry of first row and second column in $\mathcal{A}_{1}$ is $\frac{b}{2}$, which is, as mentioned, is transition probability from vertex $1$ to $2$ in $K_{3}$.
The corresponding minimal idempotent of the obtaining adjacency matrices with respect to Eq. (\ref{idempotenttt}) are
\begin{equation}
S_{1(2)}(0)=\frac{1}{3} \left(
                   \begin{array}{cccc}
                     1 & 1 & 1 \\
                     1 & 1 & 1 \\
                     1 & 1 & 1 \\
                   \end{array}
                 \right), \quad
S_{1(2)}(1)=\frac{1}{3} \left(
                   \begin{array}{cccc}
                      2 &   -1   &  -1 \\
                      -1 &    2   &  -1  \\
                     -1 &   -1   &   2  \\
                   \end{array}
                 \right)
\end{equation}
In last step, with respect to Eq. (\ref{mainly}), eigenvalues of $A_{1}(\alpha_{1})$ ($A_{2}(\alpha_{2})$) on $S_{1}(\alpha_{1})$ ($S_{2}(\alpha_{2})$) are
$$
P_{1}(0,0)=P_{2}(0,0)=1, \quad P_{1}(0,1)=P_{2}(0,1)=1
$$
\begin{equation}
P_{1}(1,0)=P_{2}(1,0)=2, \quad P_{1}(1,1)=P_{2}(1,1)=-1
\end{equation}
Finally, we have
$$
\frac{2}{3}\left(1-e^{-\lambda}\right)\leq \frac{\rho_{L}D}{\frac{\rho_{E}^{2}D^{2}}{W_{E}^{\frac{2}{3}}}-2\rho_{L}^{2}-\frac{\rho_{L}\rho_{E}D}{W_{E}^{\frac{1}{3}}}} \leq \frac{2}{3}\left(1+e^{-\lambda}\right)$$
\begin{equation}\label{region}
\frac{2}{3}\left(1-e^{-\lambda}\right)\leq \frac{\rho_{L}D'}{\frac{\rho_{E}^{'2}D^{'2}}{W_{E}^{'\frac{2}{3}}}-2\rho_{L}^{2}-\frac{\rho_{L}\rho_{E}'D'}{W_{E}^{'\frac{1}{3}}}} \leq \frac{2}{3}\left(1+e^{-\lambda}\right)
\end{equation}
which $\lambda$ is the Lyapunov exponent of the isolated UCAs. By applying the direct product of two association schemes, the stability condition of the composed cluster of the albumin and polymer agents (see Fig. 3), can be written as follows:
\begin{equation}\label{froha}
\frac{1}{3}\left(1-e^{-\lambda}\right)\leq  \Delta \leq \frac{1}{3}\left(1+e^{-\lambda}\right)
\end{equation}
where
$$
\Delta =\frac{\rho_{L}D}{\frac{\rho_{E}^{2}D^{2}}{W_{E}^{\frac{2}{3}}}-2\rho_{L}^{2}-\frac{\rho_{L}\rho_{E}D}{W_{E}^{\frac{1}{3}}}}+\frac{\rho_{L}D'}{\frac{\rho_{E}^{'2}D^{'2}}{W_{E}^{'\frac{2}{3}}}-2\rho_{L}^{2}-\frac{\rho_{L}\rho_{E}'D'}{W_{E}^{'\frac{1}{3}}}}-\frac{3\rho_{L}^{2}D  D'}{4\left[\frac{\rho_{E}^{2}D^{2}}{W_{E}^{\frac{2}{3}}}-2\rho_{L}^{2}-\frac{\rho_{L}\rho_{E}D}{W_{E}^{\frac{1}{3}}}\right]\left[\frac{\rho_{E}^{'2}D^{'2}}{W_{E}^{'\frac{2}{3}}}
-2\rho_{L}^{2}-\frac{\rho_{L}\rho_{E}'D'}{W_{E}^{'\frac{1}{3}}}\right]}
$$
and details of the  direct product of two clusters is presented in \ref{productcluste}. Clearly, by combining two different clusters, their stability condition becomes more complicated than one cluster.

The stability condition at synchronization state is completely dependent on the dynamical characteristics of single microbubble which is denoted by the Lyapunov exponent. For the cluster with $N$ element, it is simple to generate the general form of Eq. (\ref{lyappa}). In order to detect the stable domain at synchronized state, the study of $N$ interacting microbubble cluster  is reduced to study of the isolated case.
\section{Results}\label{Sec4}
Since three interacting microbubbles have two associated classes $\beta=0$  and $\beta=1$, so by applying $S(0)$ and $S(1)$ on the UCAs cluster phase space $R_{2i}(t)$, we find that their trajectories transit to two modes: synchronization state by means of $S(0)$  which identify the isolated microbubble synchronization state and transversally mode due to $S(1)$ which represents synchronization state between three microbubbles. On the other hand, in the transversal state, microbubble cluster has the stable or chaotic attractors which depend on $S(1)$, the microbubbles present in the stable or chaotic synchronous mode.

Eq. (\ref{region}) shows that the stability regions of UCAs interaction with different agents at CS state is dependent on the distance between targeted microbubbles $(D)$, shell thickness $(R_{20}-R_{10})$ with respect to $(W_{E})$, shell density $(\rho_{E})$, liquid density $(\rho_{L})$ and the isolated UCAs Lyapunov exponent $(\lambda)$. On the other hand, the characteristic dynamical behavior of single targeted microbubble is dependent on several control parameters, such as acoustic pressure ($P_{ac}$), ultrasound frequency ($\nu$), shell viscosity ($\mu_{E}$), and shell thickness ($R_{20}-R_{10}$). On the other hand, for permissible values of the distance between microbubbles, the UCAs cluster can become at CS state, which is obtained analytically by using $\lambda$ versus control parameters of single UCAs. It should be noted that $\lambda$ has three values, when $\lambda<0$, single targeted microbubble is in the stability region and it represents periodic behavior, if $\lambda=0$, the system exhibits quasi-periodic behavior which shows microbubble tends to transit to the chaotic region. For $\lambda>0$, single encapsulated microbubble represents its chaotic nature. Clearly, depending on these  three values, obtained inequality condition for microbubble network at synchronous mode demonstrate different behavior, whereby considering term $e^{-\lambda}$, stable or quasi-periodic states have larger synchronization condition than the chaotic state. Moreover, by investigation of dynamical behaviors of single targeted microbubble by using the   Lyapunov exponent diagrams,  the most effective parameter in order to control the interacting microbubble cluster becomes available. In the following, by plotting Lyapunov exponent diagrams of four important parameters, their intrinsic behaviors reveal the stable synchronization state. The stable states of two different agents are presented in Table 2.

\subsection{Effect of acoustic pressure}
The amplitude of acoustic pressure is one of the most important external parameters on UCAs cluster dynamics which has a significant role in  determing   the stability region.  The maximum Lyapunov exponent of versus acoustic pressure (varies from $10$ KPa to $2$ MPa) for albumin and polymer agents of the isolated UCAs is plotted in Fig. 4. The maximum Lyapunov exponent is an important indicator for a dynamic system for detecting potentially chaotic behavior. The results indicate that the chaotic oscillations and the expansion ratio of UCAs have a direct relationship with the magnitudes of acoustic pressure which is seen in~\cite{sarkar,marmo}. Where the Lyapunov exponent is mostly positive which shows a chaotic
behavior. The negative Lyapunov exponent demonstrates the stable behavior.  {From the results, it can be understood that the motions of the microbubble can be chaotic or stable in particular ranges.}

The UCA with an albumin layer (Fig. 4(a)) reveals high nonlinearities and high expansion ratio when it is stimulated with high acoustic pressure amplitudes, i.e., about $1.2$ MPa which this incident occurred in $0.95$ MPa for polymer (Fig. 4(b)). { Also, the corresponding bifurcation diagram of the normalized albumin-agent microbubble radius are shown in Fig. 5. The chaotic behavior of the albumin-agent appears by increasing the values of applied pressure and the microbubble shows more chaotic oscillations as the pressure is intensifying and this phenomenon is seen in Keller-Herring model for UCA \cite{macdonbnh} and the Keller-Miksis model \cite{Garashchuk18}.} So the transition time from the stable to chaotic states for the albumin agent become longer than polymer case which shows the stable regions of the microbubble cluster for albumin agent are wider than polymer shell. On the other hand, by gradually increasing of the amplitude of acoustic pressure, microbubbles lead towards chaotic synchronization state. The UCAs behavior is stable for low values of acoustic pressure and chaotic oscillations are inevitable in severe acoustic pressures.
\subsection{Effect of ultrasound frequency}
One of the most influential parameters that affect the microbubble dynamics is the applied ultrasound frequency which is utilized as a control parameter and varying in the range of $0.8$ to $5$ MHz, where $P_{ac}$=1.5 MPa (see Fig. 6).

For albumin agent (Fig. (6)) the radial motion of the microbubble is completely chaotic and large expansion amplitudes are apparent for low values of frequency up to $1$ MHz, however by increasing the ultrasound frequency the chaotic pulsations and the expansion ratio of the UCA is reduced significantly and the system becomes totally stable via period one in $1.5$ MHz. For example, for polymer-agent, the chaotic region and the microbubble expansion ratio becomes smaller by increasing the frequency values and the UCA demonstrates the stable behavior of period one in $2$ MHz. It is evident that the microbubble is stable in high amplitudes of frequency for two agents and these figures disclose the stabilizing effect of superior frequencies which is also reported by~\cite{allen}. { It is seen in the figure 7 that, the magnitude
of pulsations reduces significantly and chaotic region becomes smaller when the
 control parameter (frequency) is increasing and the albumin-shelled agent shows stable behavior
of period one which reveals the stabilizing property of superior frequencies which
is confirmed in \cite{macdonbnh,behnia2,Garashchuk18}.} Moreover, one can observes a reverse period-doubling cascade ({period doubling})~\cite{parliz} which can be found in periodic orbit. By considering Fig. 4 and Fig. 6, it can be observed that for the albumin agent with $ \nu \simeq 3$ MHz and polymer-agent with $\nu \simeq 2$ MHz single microbubble exhibit a cycle of periodic with subsequent bifurcations. So these  fixed points unstable with $\nu > 3$ MHz for the albumin and $\nu > 2$ MHz for polymer agents.
\subsection{Effect of shell viscosity}
Lyapunov exponent diagram of the normalized microbubble radius in Fig. 8 is demonstrated by taking the shell viscosity as the control parameter whereas the acoustic pressure amplitude is $1.5$ MPa and the ultrasound frequency is $1$ MHz. The effect of shell viscosity variations on UCAs behavior is plotted from $1$ to $10$ Pa s. As can be observed in Fig. 8, the maximum expansion ratio of albumin-shelled agent is high for small values of shell viscosity and its nonlinear oscillations abate as the shell viscosity is increasing.  {Figure 9 shows the bifurcation diagrams of the normalized albumin-agent microbubble radius when shell viscosity of the microbubble is taken as the control parameter with the $P_{ac}$=1.5 MPa and $\nu$=1 MHz, whose chaotic stable and  pulsations can be observed in respective parts of figure.} This agent becomes stable with period one in $4$ Pa s. Albumin and polymer shelled agents obey relatively the same trend when the shell viscosity is increasing. Their dynamical behaviors started with chaotic behavior which by increasing shell viscosity they lead towards the stable region. Then by increasing amounts of viscosity, interacting targeted microbubble cluster at CS state present in the stable region.

\subsection{Effect of shell thickness}
The dynamics of acoustically stimulated UCAs is inspected by applying the shell thickness of the agents as the control parameter while altering between $0$ to $150$ $nm$ in the acoustic pressure $1.5$ MPa and ultrasound frequency $1$ MHz. The dynamical response of the agents versus variations of the shell thickness is shown in Fig. 10. The microbubble exhibits fully chaotic behavior for small values of the shell thickness.  The results disclose that nonlinear oscillations decreases and the stability of the agents increases as their shell thickness is raising which are in agreement with the previous works~\cite{allen,khismatullin}.
Fig. 10(a)  {(also bifurcation diagram of figure 11)} belongs to an albumin-shelled agent which indicates a considerable reduction in expansion ratio of the microbubble by increasing its shell thickness. It has experiences the stability of period one in the amplitude for the shell thickness equal to $30$ nm. However, for the polymer shell (see Fig. 10(b)), the result shows, it experiences the stability of period one in the amplitude of the shell thickness equal to $120$ nm. So there is a big difference between two different agents in the synchronization condition.
\section{Conclusion}\label{Sec5}
In summary, the dynamical behaviors of $N$ interacting targeted microbubble cluster under the action of a high acoustical pressure at complete synchronization state are investigated in the framework of association schemes theory and Bose-Mesner algebra. The applied procedure can be facilitated the understanding of cluster formation from the ultrasound echoes and the stable behavior of UCAs network at synchronous mode. The dynamical properties of the isolated targeted microbubble have the crucial role in the stability behavior of \textbf{a very large} number of interacting encapsulated microbubble cluster which is immersed into the blood or soft tissue.  {On the other hand, several mathematical models, identify bubbles oscillations in a cluster, such as linear theory \cite{Ida02,Zhang97} or self-consistent oscillator model \cite{Feuillade01,Feuillade95,Ye97}. When the number of bubbles in a cluster is increased, a signiﬁcant error between experimental and theoretical results appear.} When single UCAs in an ultrasound field has chaotic behavior $\lambda>0$, we receive the acceptable bounds on stable synchronous mode which become smaller than the stable or quasi-periodic case.
Since targeted microbubbles in a high acoustical pressure with nonlinearity behavior are at chaotic state, this procedure shows that nanoscale dispersed microbubbles whic satisfy Markovian matrices property and projecting of minimal idempotent $S(\beta)$ on space states tends to oscillate simultaneously. The distance between interacting UCAs cluster has the main role in their synchronization states, which shows the influence of coupling between microbubbles is always significant and is no longer negligible~\cite{dzah}. While coated microbubbles bound to targeted surfaces, their clustering may occur by means of binding drugs to a group of microbubbles shell and fragmentation of targeted microbubbles by an ultrasound insonification, the localized release of the drugs perform in the region interest~\cite{rapopa}. On the other hand, the results show,  distances, the properties of agents, shell density and thickness are effective in dynamical behaviors of multiple microbubbles which are already was predicted by Allen et. al~\cite{alllem}. So a well-designed group of targeted microbubbles, in order to pharmacotherapy applications, is possible by using of obtained synchronization condition for UCAs cluster in the framework of Bose-Mesner algebra.

A typical 5-$cc$ injection of $2.5$-micron agents into (human) blood with mixing homogeneously depends on the concentration contains approximately $1-5\times10^{11}$ agents~\cite{alllem}. So the analysis of these larger microbubble clusters with different agents and distances is approximately impossible. However, by using a combination of two different association schemes, we show that a composed microbubbles network of different agents can be presented at synchronized states which represented their deterministic chaotic state. As can be seen, their stable behavior is more and more complicated than simple networks. Therefore, by introducing direct product of two different schemes, we can be observed as an extension of interacting targeted microbubbles to network motifs from smaller ones. {Based on the obtained results it can be concluded that the significance of this paper is that the stability of the synchronized state (or symmetric eigenmode of mutual bubble oscillation) with respect to another state (another eigenmode) can now be predicted.} Finally, this procedure may help us to control the nanoscale dispersed coupled targeted microbubbles for diagnostic due to scattering UCAs and therapeutic applications.

\section*{Acknowledgement}
Author M.Y would like to thanks Luqman Saleem for helping in editing the paper.
\appendix{

\section{Association scheme}\label{AppendixA}
A graph is a pair $(V,E)$ , where $V$ is vertices set and an edge set $E$ is an unordered pair of distinct vertices. Two vertices $i,j \in V$ are called adjacent if $\{i, j\}\in E$. The adjacency matrix of $(V,E)$, which rows and columns corresponding to the vertices $i,j$ is defined by~\cite{grossk}
$$
(A)_{i,j}=\left\{\begin{array}{cc}
                    1 & if \;\; i \sim j \\\\
                     0 & otherwise
                   \end{array}\right.
$$
obviously, $A$ is a symmetric matrix. The valency of a vertex, $i \in V$ is defined as
$$\kappa(i)=\mid\{j \in V : i \sim j \}\mid$$
where $\mid.\mid$ denotes the measure of the number of elements of the $V$. Let $V$ be a set of vertices and $\mathcal{B}(\alpha)=\{\mathcal{B}(0), \mathcal{B}(1), \cdot\cdot\cdot, \mathcal{B}(d)\}$  a nonempty set of relationships on $V$ which is called associate class. The pair $\tilde{X}=(V,\mathcal{B}(\alpha))$ which satisfies four conditions, is an association scheme of class $d$ ($d$-class scheme) on $V$~\cite{bailey,brouw}.
The $d+1$ associate class $\mathcal{B}(\alpha)$ of an association scheme is described by their adjacency matrices $A(\alpha)$, defined by $A(\alpha)_{i,j} = 1$ if $(i,j) \in \mathcal{B}(\alpha)$ ; 0 otherwise. The $A(\alpha)$ are the adjacency matrices of $\tilde{X}$, where the other definition of an association scheme is equivalent to the following~\cite{brouw,bannai}
$$
\sum_{\alpha=0}^{d} A(\alpha)=J_{N},\quad
A(0)=I_{N}$$
\begin{equation}\label{matr}
A(\alpha)=A^{T}(\alpha) ,\quad
A(\alpha)A(\beta)=\sum_{\gamma=0}^{d} p_{\alpha \beta}^{\gamma} A(\gamma), \quad 0 \leq \alpha,\beta \leq d
\end{equation}
the numbers $p_{\alpha \beta}^{\gamma}$, are called the intersection numbers of the association scheme~\cite{bailey,bos}. $J$ is a $N\times N$ matrix with all-one entries. The relationship between the number of vertices (or order of association scheme) and valency is given by:
\begin{equation}\label{sareza}
N=|V|=\sum_{\alpha=0}^{d}\kappa(\alpha)
\end{equation}
where
\begin{equation}\label{valencyy}
\kappa(\alpha)=p_{\alpha \alpha}^{0},\quad \kappa(\alpha) \neq 0
\end{equation}
$A(0), A(1), \cdot \cdot \cdot, A(d)$  generate a commutative $(d+1)$-dimensional algebra $\mathbf{A}$ of symmetric matrices which called Bose-Mesner algebra of $\tilde{X}$~\cite{bose}.
On the other hand, $\mathbf{A}$ has second basis $S(0),..., S(d)$ ~\cite{bailey}, satisfy
\begin{equation}\label{idempotenttt}
S(0) = \frac{1}{N}J_{N},\quad S(\alpha)
S(\beta)=\delta_{\alpha \beta} S(\alpha),\quad \sum_{\alpha=0}^d
S(\alpha)=I_{N}, \quad 0\leq \alpha,\beta \leq d
\end{equation}
{for $\alpha$ or $\beta$ $= 0,\;1,\;\cdots,\;d$, the stratum projector (the stratum projectors are sometimes called  primitive idempotents ~\cite{bailey}) $S(\alpha)$ or $S(\beta)$ is a linear combination of $A_0$, $A_1$, $\cdots$, $A_d$.} Actually, the $S(\alpha), S(\beta)$ are known as the primitive idempotent of $\tilde{X}$, where the matrix $\frac{1}{N}J_{N}$ is a minimal idempotent. If $P$ and $\frac{1}{N}Q$ be the matrices relating our two bases for $\mathbf{A}$, adjacency matrix expand with respect to the primitive idempotent as follows~\cite{bannai,godsill}
\begin{equation}\label{eigen1}
A(\alpha)=\sum_{\beta=0}^d P(\alpha,\beta)S(\beta),
\quad
S(\beta)=\frac{1}{N}\sum_{\alpha=0}^d Q(\alpha,\beta)A(\alpha),
 \;\;\;\;\ 0\leq
\alpha,\beta\leq d
\end{equation}
On the other hand, the matrices $P$ and $Q$, clearly follows from $PQ=QP=NI_{N}$, then
\begin{equation}\label{mainly}
A(\alpha) S(\beta)=P(\alpha,\beta)S(\beta)
\end{equation}
where $P(\alpha,\beta)$, $Q(\alpha,\beta)$ is the $\alpha$-th eigenvalues of $A(\alpha)$, $S(\alpha)$, respectively.
The other parameter, which is the relationship between two eigenvalues is $m(\alpha)$, can be defined as
$$
m(\alpha)=Tr\left(S(\alpha)\right),\quad
\sum_{\alpha=0}^{d}m(\alpha)=N
$$
\begin{equation}
m(\beta) P(\alpha,\beta)=\kappa(\alpha) Q(\alpha,\beta),\quad
m(0)=1, \quad 0\leq
\alpha,\beta \leq d
\end{equation}
\section{Direct product of association schemes}\label{AppendixB}
The product of schemes is important not only for constructing new association schemes from old ones, but also for describing the structure of certain schemes. The direct product of association schemes was first introduced by Kusumoto~\cite{kusan}. Let $\mathcal{B}_{1}(\alpha_{1})$ have $d_{1}$ associate classes with valencies $\kappa_{1}(\alpha_{1})$ and adjacency matrices $A_{1}(\alpha_{1})$ with eigenvalues $P_{1}(\alpha_{1},\beta_{1})$ ($0\leq \alpha_{1}, \beta_{1} \leq d_{1}$). Similarly, $\mathcal{B}_{2}(\alpha_{2})$ with $d_{2}$ associate classes, valencies $\kappa_{2}(\alpha_{2})$ have adjacency matrices $A_{2}(\alpha_{2})$ with eigenvalues $P_{2}(\alpha_{2},\beta_{2})$ ($0 \leq \alpha_{2}, \beta_{2} \leq d_{2}$). Then, $\mathcal{B}(\alpha_{1}) \times \mathcal{B}(\alpha_{2})$ is an association scheme on $V_{1} \times V_{2}$ with $(d_{1}d_{2}+d_{1}+d_{2})$ associated classes and valencies~\cite{bailey}
\begin{equation}\label{kappa}
\kappa_{1}(\alpha_{1})\kappa_{2}(\alpha_{2})
\end{equation}
The direct product of two adjacency matrices represented by:
\begin{equation}\label{zarb}
A_{1}(\alpha_{1}) \otimes A_{2}(\alpha_{2})
=\left(
\begin{array}{ccc}
  a_{1,1} A_{2}(\alpha_{2})&  \cdots & a_{1,m} A_{2}(\alpha_{2}) \\
   a_{2,1} A_{2}(\alpha_{2}) &  \cdots & a_{2,m} A_{2}(\alpha_{2}) \\
  \vdots  & \vdots & \vdots  \\
  a_{n,1} A_{2}(\alpha_{2}) &  \cdots & a_{n,m} A_{2}(\alpha_{2}) \\
 \end{array}
 \right)
\end{equation}
where $n$ and $m$ is the row and column of $A(\alpha_{1})$ and ''$\otimes$'' denotes the Kronecker product. The direct product of two minimal idempotent likewise adjacency matrices is defined as
$$
S_{1}(\alpha_{1}) \otimes S_{2}(\alpha_{2})
$$
Finally, the eigenvalues of $A_{1}(\alpha_{1}) \otimes A_{2}(\alpha_{2})$ on $S_{1}(\alpha_{1}) \otimes S_{2}(\alpha_{2})$ is given by
\begin{equation}\label{eigenb}
P_{1}(\alpha_{1},\beta_{1}) P_{2}(\alpha_{2},\beta_{2})
\end{equation}

\section{Direct product of two different cluster}
\label{productcluste}
By considering two obtained coupling matrices in Eq. (\ref{markovooo}), the coupling matrix of constructing cluster with respect to Eq. (\ref{zarb}) is given by
$$
\mathcal{C}= aa' A_{1}(0)\otimes A_{2}(0)+ \frac{ab'}{2} A_{1}(0)\otimes A_{2}(1)+ \frac{ba'}{2}A_{1}(1)\otimes A_{2}(0)+\frac{bb'}{4}A_{1}(1)\otimes A_{2}(1)
$$
where
$$
A_{1}(0)\otimes A_{2}(0)= \left(
                                  \begin{array}{ccccccccc}
                                    1 & 0 & 0 & 0 & 0 & 0 & 0 & 0 & 0 \\
                                    0 & 1 & 0 & 0 & 0 & 0 & 0 & 0 & 0 \\
                                    0 & 0 & 1 & 0 & 0 & 0 & 0 & 0 & 0 \\
                                    0 & 0 & 0 & 1 & 0 & 0 & 0 & 0 & 0 \\
                                    0 & 0 & 0 & 0 & 1 & 0 & 0 & 0 & 0 \\
                                    0 & 0 & 0 & 0 & 0 & 1 & 0 & 0 & 0 \\
                                    0 & 0 & 0 & 0 & 0 & 0 & 1 & 0 & 0 \\
                                    0 & 0 & 0 & 0 & 0 & 0 & 0 & 1 & 0 \\
                                    0 & 0 & 0 & 0 & 0 & 0 & 0 & 0 & 1 \\
                                  \end{array}
                                \right), \quad
A_{1}(0)\otimes A_{2}(1)=\left(
                      \begin{array}{ccccccccc}
                        0 & 1 & 1 & 0 & 0 & 0 & 0 & 0 & 0 \\
                        1 & 0 & 1 & 0 & 0 & 0 & 0 & 0 & 0 \\
                        1 & 1 & 0 & 0 & 0 & 0 & 0 & 0 & 0 \\
                        0 & 0 & 0 & 0 & 1 & 1 & 0 & 0 & 0 \\
                        0 & 0 & 0 & 1 & 0 & 1 & 0 & 0 & 0 \\
                        0 & 0 & 0 & 1 & 1 & 0 & 0 & 0 & 0 \\
                        0 & 0 & 0 & 0 & 0 & 0 & 0 & 1 & 1 \\
                        0 & 0 & 0 & 0 & 0 & 0 & 1 & 0 & 1 \\
                        0 & 0 & 0 & 0 & 0 & 0 & 1 & 1 & 0 \\
                      \end{array}
                    \right)$$

$$
A_{1}(1)\otimes A_{2}(0)=\left(
                        \begin{array}{ccccccccc}
                          0 & 0 & 0 & 1 & 0 & 0 & 1 & 0 & 0 \\
                          0 & 0 & 0 & 0 & 1 & 0 & 0 & 1 & 0 \\
                          0 & 0 & 0 & 0 & 0 & 1 & 0 & 0 & 1 \\
                          1 & 0 & 0 & 0 & 0 & 0 & 1 & 0 & 0 \\
                          0 & 1 & 0 & 0 & 0 & 0 & 0 & 1 & 0 \\
                          0 & 0 & 1 & 0 & 0 & 0 & 0 & 0 & 1 \\
                          1 & 0 & 0 & 1 & 0 & 0 & 0 & 0 & 0 \\
                          0 & 1 & 0 & 0 & 1 & 0 & 0 & 0 & 0 \\
                          0 & 0 & 1 & 0 & 0 & 1 & 0 & 0 & 0 \\
                        \end{array}
                      \right),\quad
A_{1}(1)\otimes A_{2}(1)= \left(
                       \begin{array}{ccccccccc}
                         0 & 0 & 0 & 0 & 1 & 1 & 0 & 1 & 1 \\
                         0 & 0 & 0 & 1 & 0 & 1 & 1 & 0 & 1 \\
                         0 & 0 & 0 & 1 & 1 & 0 & 1 & 1 & 0 \\
                         0 & 1 & 1 & 0 & 0 & 0 & 0 & 1 & 1 \\
                         1 & 0 & 1 & 0 & 0 & 0 & 1 & 0 & 1 \\
                         1 & 1 & 0 & 0 & 0 & 0 & 1 & 1 & 0 \\
                         0 & 1 & 1 & 0 & 1 & 1 & 0 & 0 & 0 \\
                         1 & 0 & 1 & 1 & 0 & 1 & 0 & 0 & 0 \\
                         1 & 1 & 0 & 1 & 1 & 0 & 0 & 0 & 0 \\
                       \end{array}
                     \right)
$$
which  shows that new association scheme has four associated classes $\mathcal{B}(0), \mathcal{B}(1), \mathcal{B}(2), \mathcal{B}(3)$.
Moreover, the coupling matrix could generate Markovian matrices. The valencies of each associated class with respect to Eqs. (\ref{kappa}), (\ref{madelk}) can be written as
\begin{equation}
\kappa_{1}(0)\kappa_{2}(0)=1, \quad \kappa_{1}(0)\kappa_{2}(1)=2, \quad \kappa_{1}(1)\kappa_{2}(0)=2, \quad \kappa_{1}(1)\kappa_{2}(1)=4
\end{equation}
The direct product of two minimal idempotent is
$$
S_{1}(0)\otimes S_{2}(0)= \frac{1}{9}\left(
                                  \begin{array}{ccccccccc}
                                    1 & 1 & 1 & 1 & 1 & 1 & 1 & 1 & 1 \\
                                    1 & 1 & 1 & 1 & 1 & 1 & 1 & 1 & 1 \\
                                    1 & 1 & 1 & 1 & 1 & 1 & 1 & 1 & 1 \\
                                    1 & 1 & 1 & 1 & 1 & 1 & 1 & 1 & 1 \\
                                    1 & 1 & 1 & 1 & 1 & 1 & 1 & 1 & 1 \\
                                    1 & 1 & 1 & 1 & 1 & 1 & 1 & 1 & 1 \\
                                    1 & 1 & 1 & 1 & 1 & 1 & 1 & 1 & 1 \\
                                    1 & 1 & 1 & 1 & 1 & 1 & 1 & 1 & 1 \\
                                    1 & 1 & 1 & 1 & 1 & 1 & 1 & 1 & 1 \\
                                  \end{array}
                                \right)$$
$$
S_{1}(0)\otimes S_{2}(1)=\frac{1}{9}\left(
                      \begin{array}{ccccccccc}
                        2 & -1 & -1 & 2 & -1 & -1 & 2 & -1 & -1 \\
                        -1 & 2 & -1 & -1 & 2 & -1 & -1 & 2 & -1 \\
                        -1 & -1 & 2 & -1 & -1 & 2 & -1 & -1 & 2 \\
                        2 & -1 & -1 & 2 & -1 & -1 & 2 & -1 & -1 \\
                        -1 & 2 & -1 & -1 & 2 & -1 & -1 & 2 & -1 \\
                        -1 & -1 & 2 & -1 & -1 & 2 & -1 & -1 & 2 \\
                        2 & -1 & -1 & 2 & -1 & -1 & 2 & -1 & -1 \\
                        -1 & 2 & -1 & -1 & 2 & -1 & -1 & 2 & -1 \\
                        -1 & -1 & 2 & -1 & -1 & 2 & -1 & -1 & 2 \\
                      \end{array}
                    \right)$$
$$
S_{1}(1)\otimes S_{2}(0)=\frac{1}{9}\left(
                      \begin{array}{ccccccccc}
                        2 & 2 & 2 & -1 & -1 & -1 & -1 & -1 & -1 \\
                        2 & 2 & 2 & -1 & -1 & -1 & -1 & -1 & -1 \\
                        2 & 2 & 2 & -1 & -1 & -1 & -1 & -1 & -1 \\
                        -1 & -1 & -1 & 2 & 2 & 2 & -1 & -1 & -1 \\
                        -1 & -1 & -1 & 2 & 2 & 2 & -1 & -1 & -1 \\
                        -1 & -1 & -1 & 2 & 2 & 2 & -1 & -1 & -1 \\
                        -1 & -1 & -1 & -1 & -1 & -1 & 2 & 2 & 2 \\
                        -1 & -1 & -1 & -1 & -1 & -1 & 2 & 2 & 2 \\
                        -1 & -1 & -1 & -1 & -1 & -1 & 2 & 2 & 2 \\
                      \end{array}
                    \right)$$
$$
S_{1}(1)\otimes S_{2}(1)=\frac{1}{9}\left(
                       \begin{array}{ccccccccc}
                         4 & -2 & -2 & -2 & 1 & 1 & -2 & 1 & 1 \\
                         -2 & 4 & -2 & 1 & -2 & 1 & 1 & -2 & 1 \\
                         -2 & -2 & 4 & 1 & 1 & -2 & 1 & 1 & -2 \\
                         -2 & 1 & 1 & 4 & -2 & -2 & -2 & 1 & 1 \\
                         1 & -2 & 1 & -2 & 4 & -2 & 1 & -2 & 1 \\
                         1 & 1 & -2 & -2 & -2 & 4 & 1 & 1 & -2 \\
                         -2 & 1 & 1 & -2 & 1 & 1 & 4 & -2 & -2 \\
                         1 & -2 & 1 & 1 & -2 & 1 & -2 & 4 & -2 \\
                         1 & 1 & -2 & 1 & 1 & -2 & -2 & -2 & 4 \\
                       \end{array}
                     \right)$$
Finally, by considering Eq. (\ref{eigenb}), the eigenvalues of $A_{1}(\alpha_{1})\otimes A_{2}(\alpha_{2})$  on $S_{1}(\alpha_{1})\otimes S_{2}(\alpha_{2})$ obtain
$$
P_{1}(0,0)P_{2}(1,0)=2, \quad P_{1}(0,0)P_{2}(1,1)=-1, \quad P_{1}(0,1)P_{2}(1,0)=2, \quad P_{1}(0,1)P_{2}(1,1)=-1
$$
$$
P_{1}(1,0)P_{2}(0,0)=2, \quad P_{1}(1,1)P_{2}(0,0)=-1, \quad P_{1}(1,0)P_{2}(0,1)=2, \quad P_{1}(1,1)P_{2}(0,1)=-1
$$
$$
P_{1}(1,0)P_{2}(1,0)=4, \quad P_{1}(1,1)P_{2}(1,0)=-2, \quad P_{1}(1,0)P_{2}(1,1)=-2$$
$$
P_{1}(0,0)P_{2}(0,0)=1, \quad P_{1}(1,1)P_{2}(1,1)=1
$$}

\clearpage
\newpage


\begin{table}
\centering
\caption{Comparison of stable regions of albumin and polymer agents.}
\label{Table2}
\begin{tabular}{|l|l|l|l|}
	\hline
Parameter (Symbol)  &Albumin    &  Polymer &Unit   \\
	\hline \hline
Viscosity ($\mu_{S} $)   &1.77  &0.45  &Pa s  \\
Density ($\rho_{S}$ )   &1100  &1150 &$\frac{kg}{m^{3}}$  \\
Shear modulus ($G_{S}$ )   &88.8  &11.7 &MPa  \\
Inner surface tension ($\sigma_{1}$)   &0.04  &0.04 &$\frac{N}{m}$  \\
Outer surface tension ($\sigma_{2}$)  &0.056  &0.056   &$\frac{N}{m}$  \\
Outer radius ($R_{20}$)   &1.5  &2.5    &$\mu$m  \\
Thickness ($R_{S}$)   &15  &125    &nm  \\
	\hline
\end{tabular}
\end{table}

\begin{table}
\centering
\caption{Comparison of stable regions of albumin and polymer agents.}
\label{Table2}
\begin{tabular}{|l|l|l|l|}
	\hline
Parameter   &Albumin    &  Polymer  &Unit  \\
   \hline \hline
Acoustic pressure    & $<$1.23  &$<$0.96     &$\textmd{MPa}$   \\
Ultrasound frequency    &$>$1.37  &$>$2     &$\textmd{MHz}$  \\
Shell viscosity   &$>$3.7   &$>$2.82     &$\textmd{Pa}\;\textmd{s}$  \\
Shell thickness    &$>$20  &$>$124     &$\textmd{nm}$   \\
	\hline
\end{tabular}
\end{table}

\begin{figure*}
\begin{center}
 \scalebox{0.54}{\includegraphics{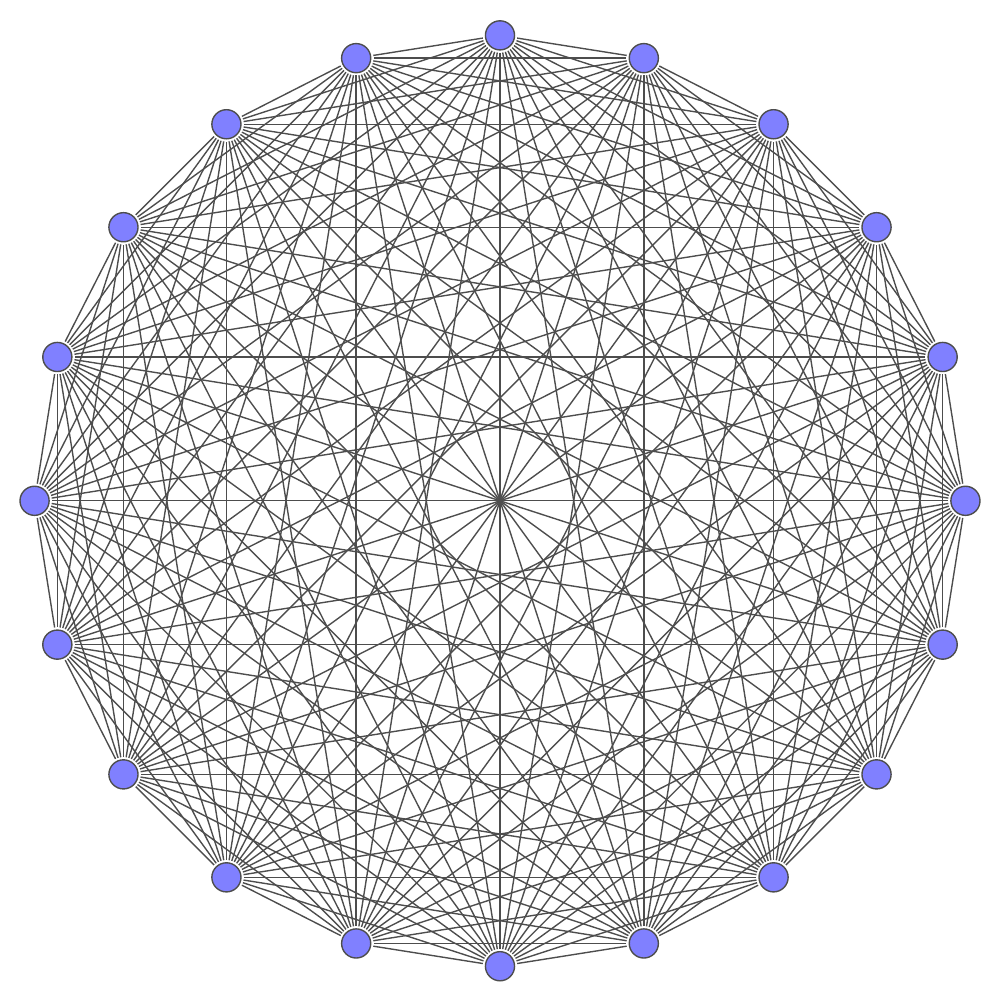}} \\
   \caption{A network of GCM ($N$=$20$), which each dynamical variable $x_{i}(t)$ is embedded into vertices of complete graph.}\label{completegraph}
    \end{center}
\end{figure*}

\begin{figure*}
\begin{center}
 \scalebox{0.54}{\includegraphics{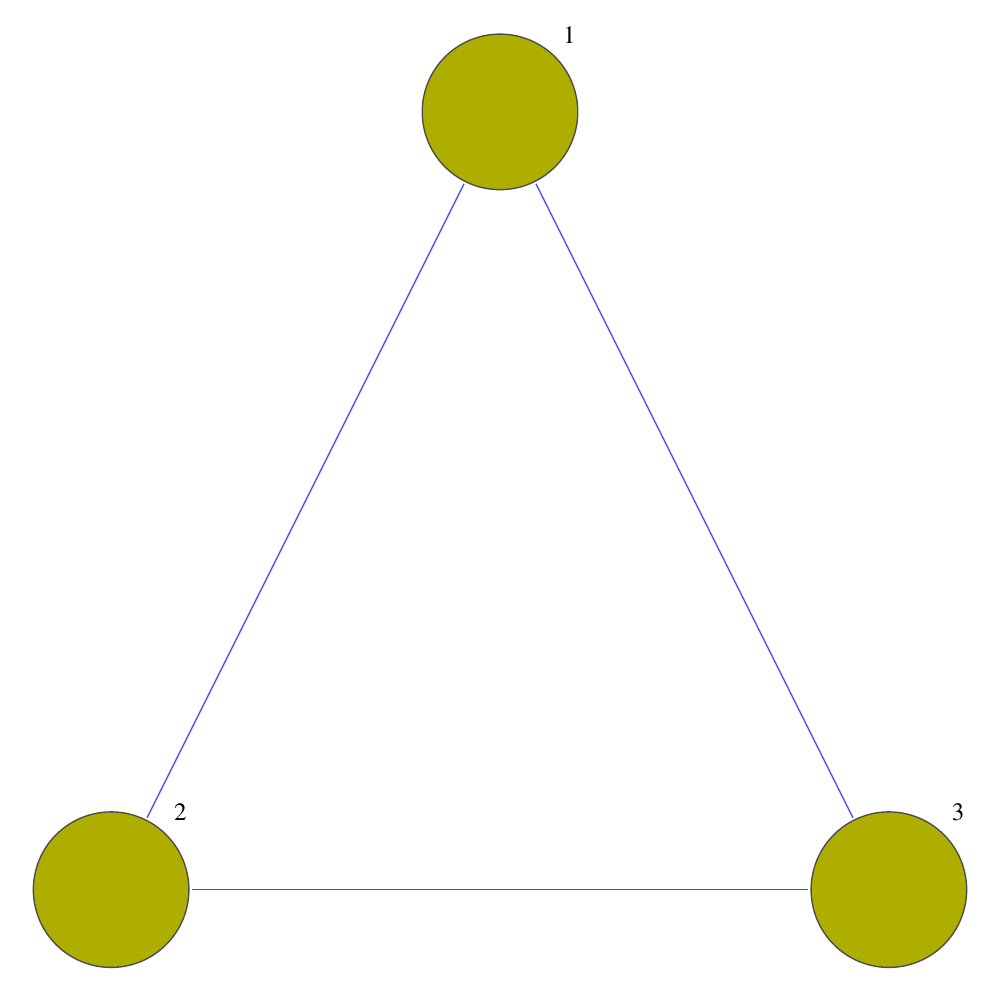}} \\
   \caption{Three interacting targeted microbubble cluster which arranged in three vertex of complete graph.}\label{threemicrobubble}
    \end{center}
\end{figure*}

\begin{figure*}
\begin{center}
 \scalebox{0.54}{\includegraphics{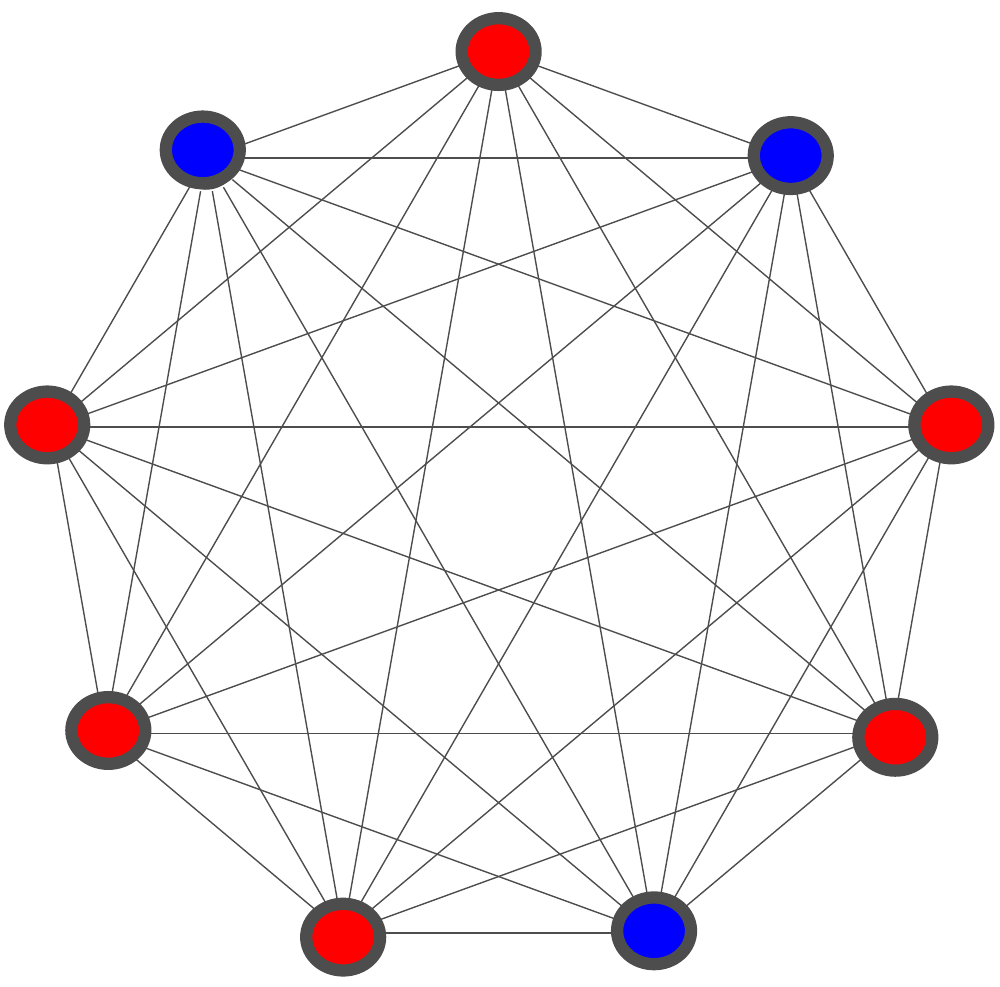}} \\
   \caption{A composed cluster of two different shell of interacting targeted microbubble which blue color is corresponding to albumin agent and red color allocate to polymer agent.}\label{network}
    \end{center}
\end{figure*}

\begin{figure*}
\begin{center}
 \scalebox{0.65}{\includegraphics{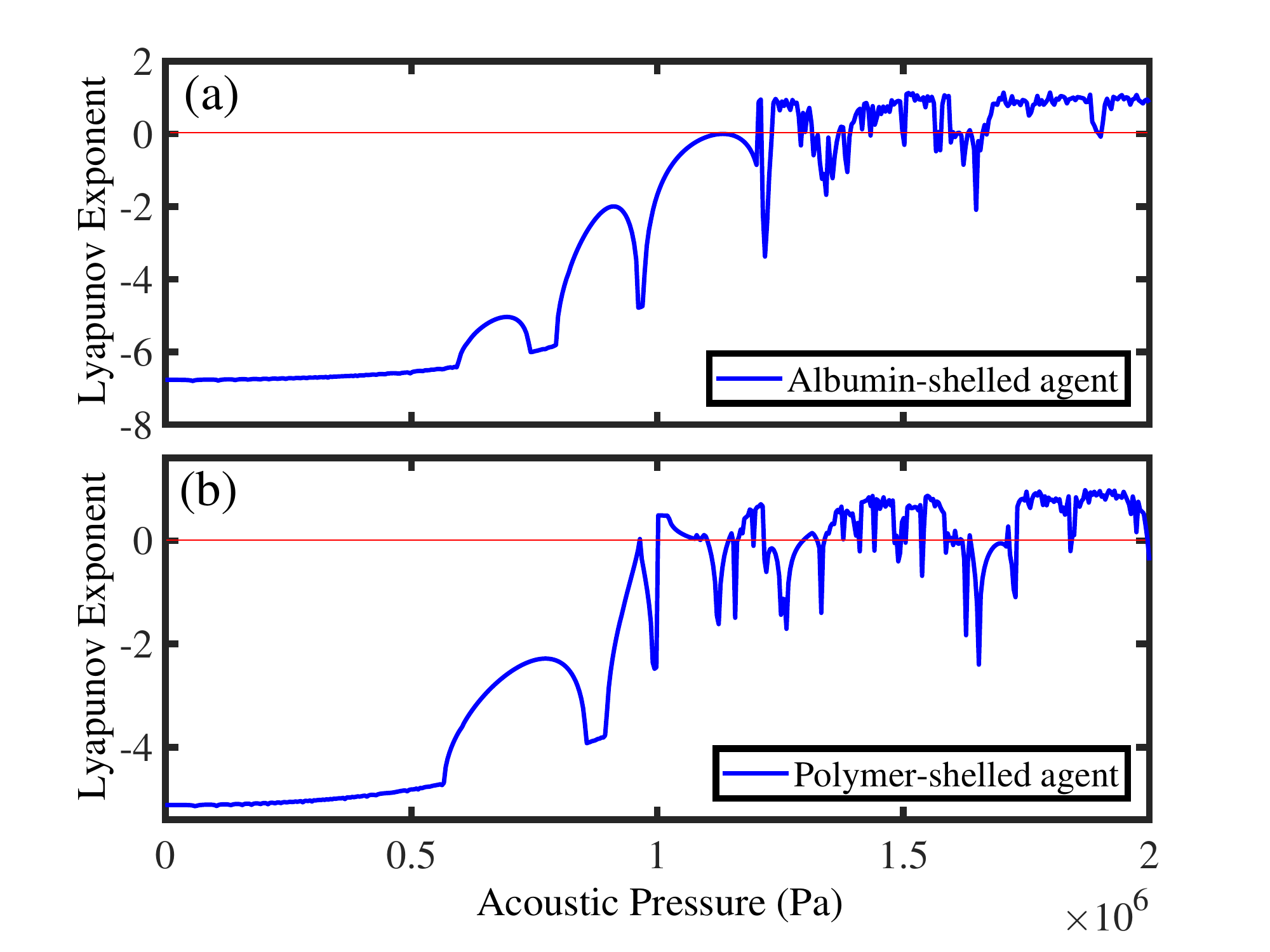}} \\
   \caption{Lyapunov spectra of normalized microbubble radius $\frac{R_{2}}{R_{20}}$ versus acoustic pressure for albumin agent  (a) and polymer agent (b) while the $\nu$=1.2 MHz.}
    \end{center}
\end{figure*}

\begin{figure*}
\begin{center}
 \scalebox{0.65}{\includegraphics{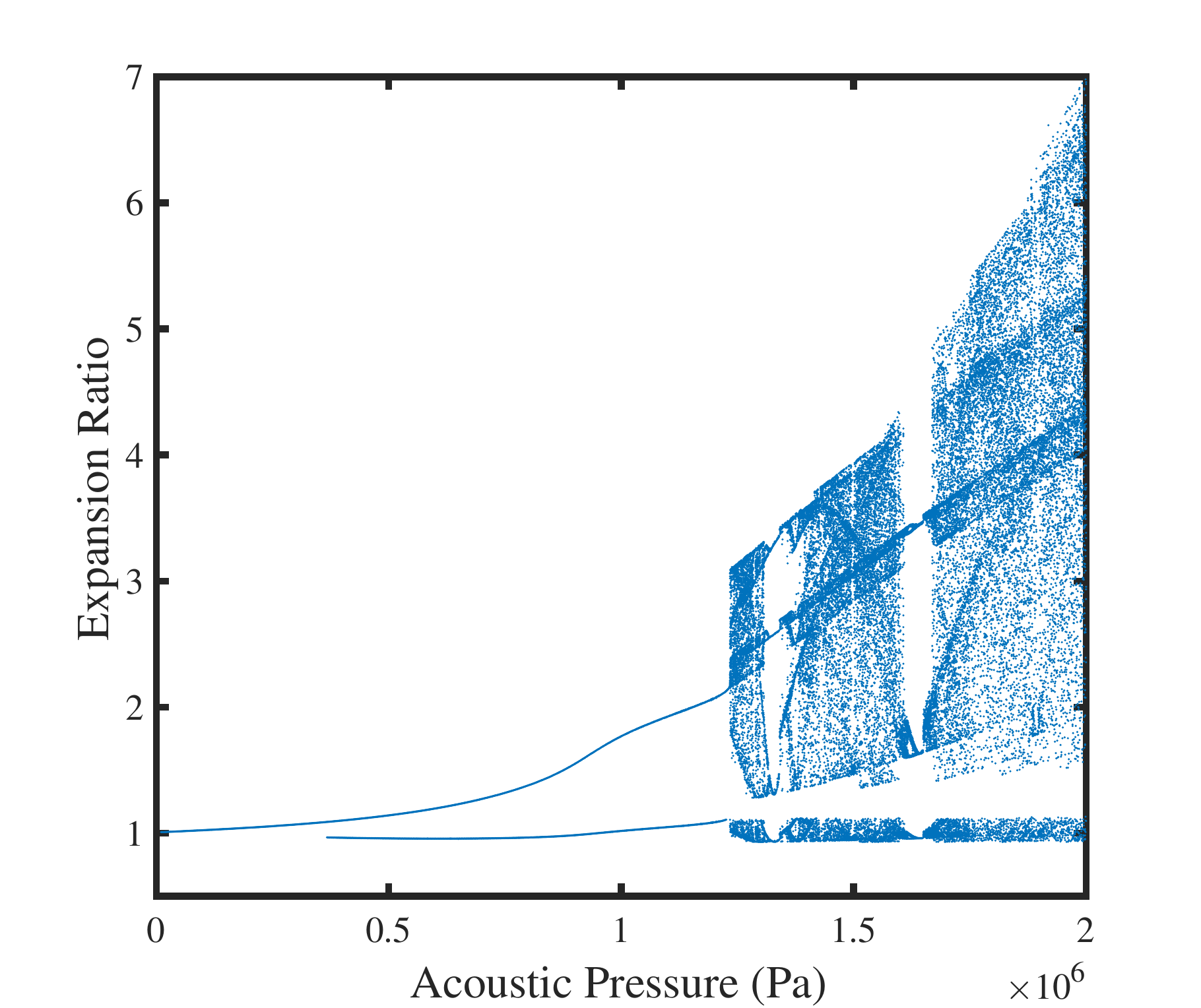}} \\
   \caption{Bifurcation diagrams of normalized albumin-agent radius versus acoustic pressure when the applied frequency is $\nu$=1.2 MHz.}
    \end{center}
\end{figure*}

\begin{figure*}
\begin{center}
 \scalebox{0.65}{\includegraphics{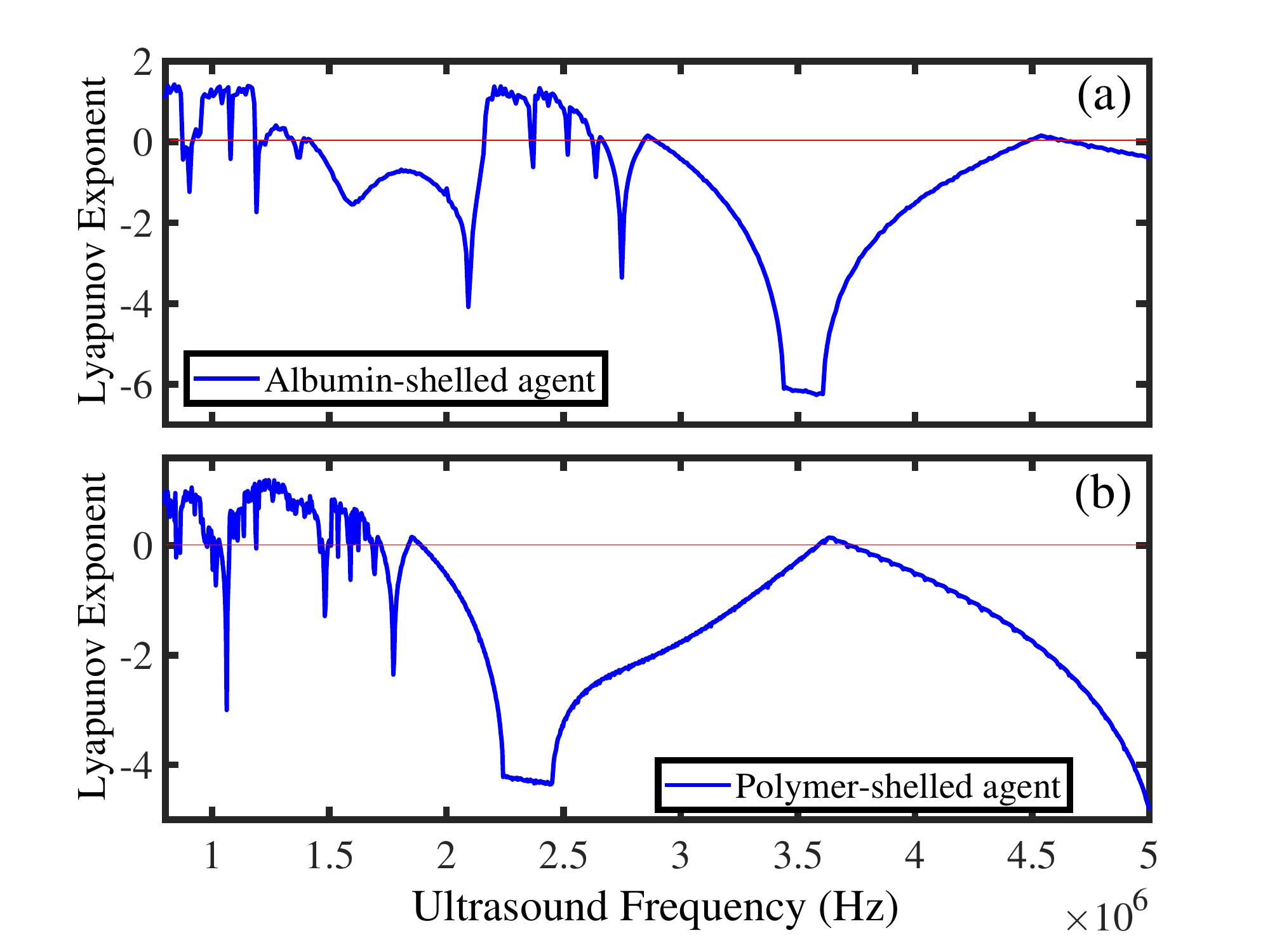}} \\
   \caption{Lyapunov spectra of normalized microbubble radius $\frac{R_{2}}{R_{20}}$ versus
ultrasound frequency for albumin agent (a) and polymer agent (b) while the $P_{ac}$=1.5 MPa.}
    \end{center}
\end{figure*}

\begin{figure*}
\begin{center}
 \scalebox{0.65}{\includegraphics{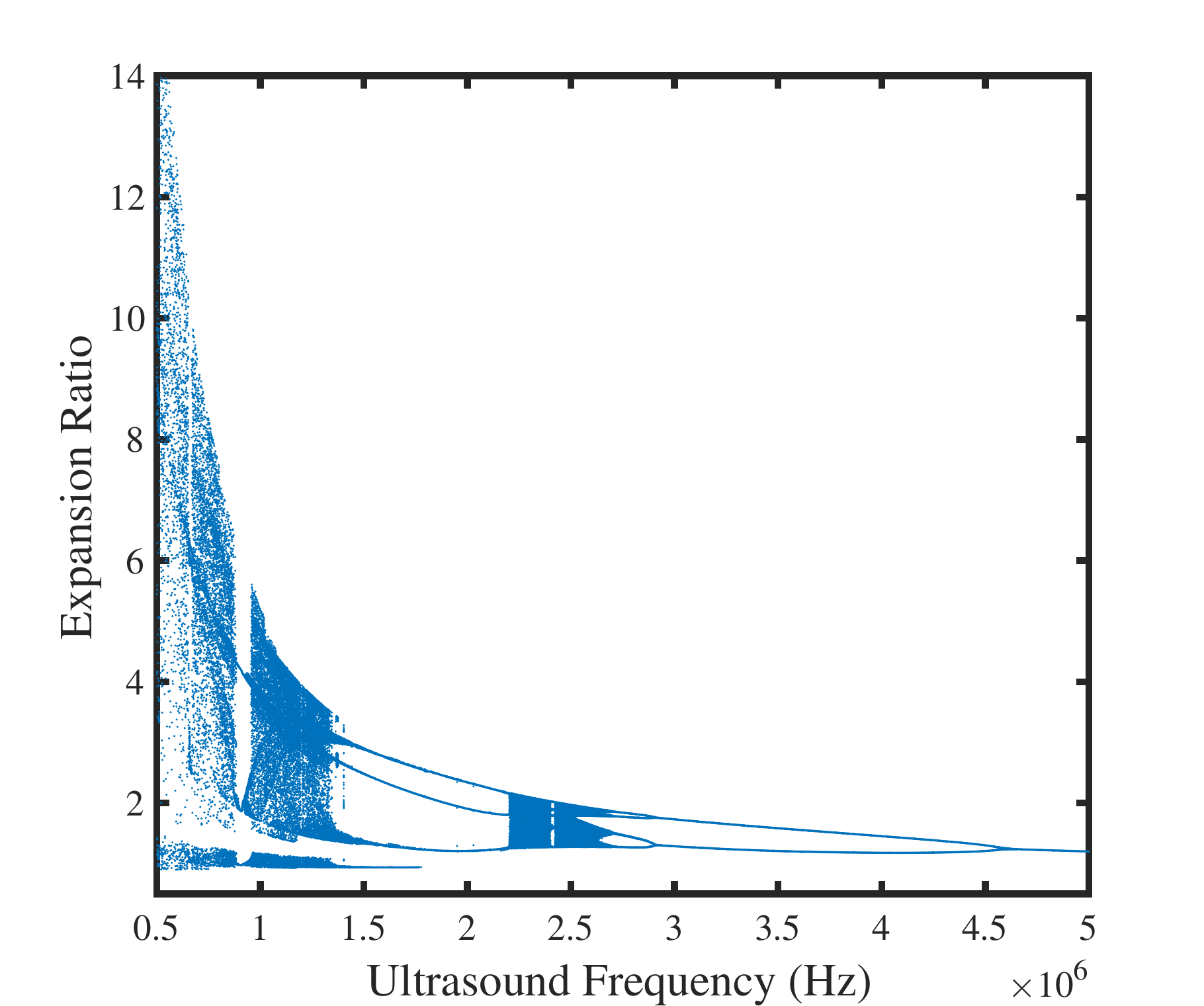}} \\
   \caption{Bifurcation diagrams of normalized albumin-agent radius versus ultrasound frequency when the acoustic pressure is $P_{ac}$=1.5 MPa.}
    \end{center}
\end{figure*}

\begin{figure*}
\begin{center}
 \scalebox{0.65}{\includegraphics{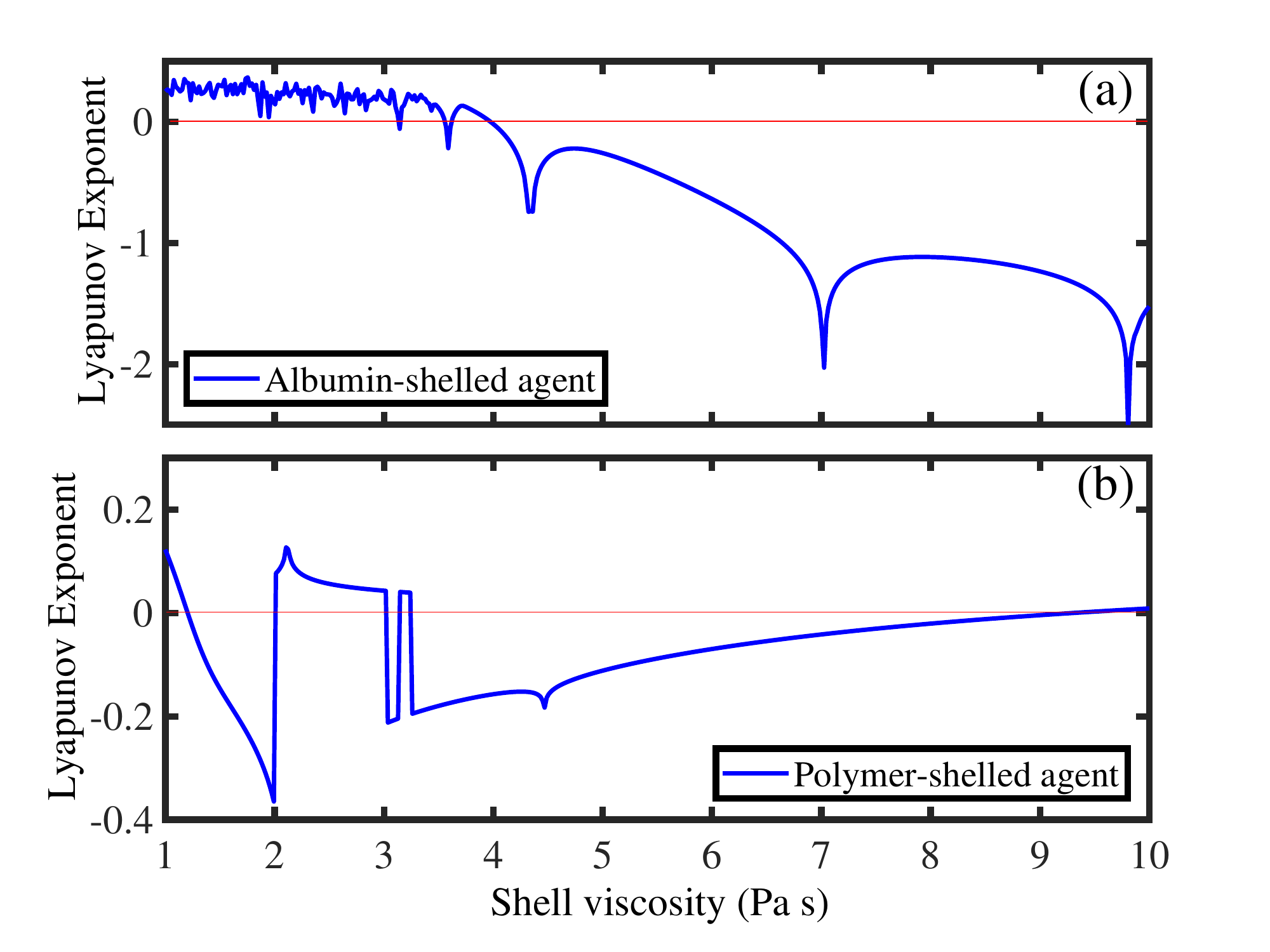}} \\
   \caption{Lyapunov spectra of normalized microbubble radius $\frac{R_{2}}{R_{20}}$ versus
shell viscosity for albumin agent (a) and polymer agent (b) while the $\nu$=1 MHz and $P_{ac}$=1.5 MPa.}
    \end{center}
\end{figure*}

\begin{figure*}
\begin{center}
 \scalebox{0.65}{\includegraphics{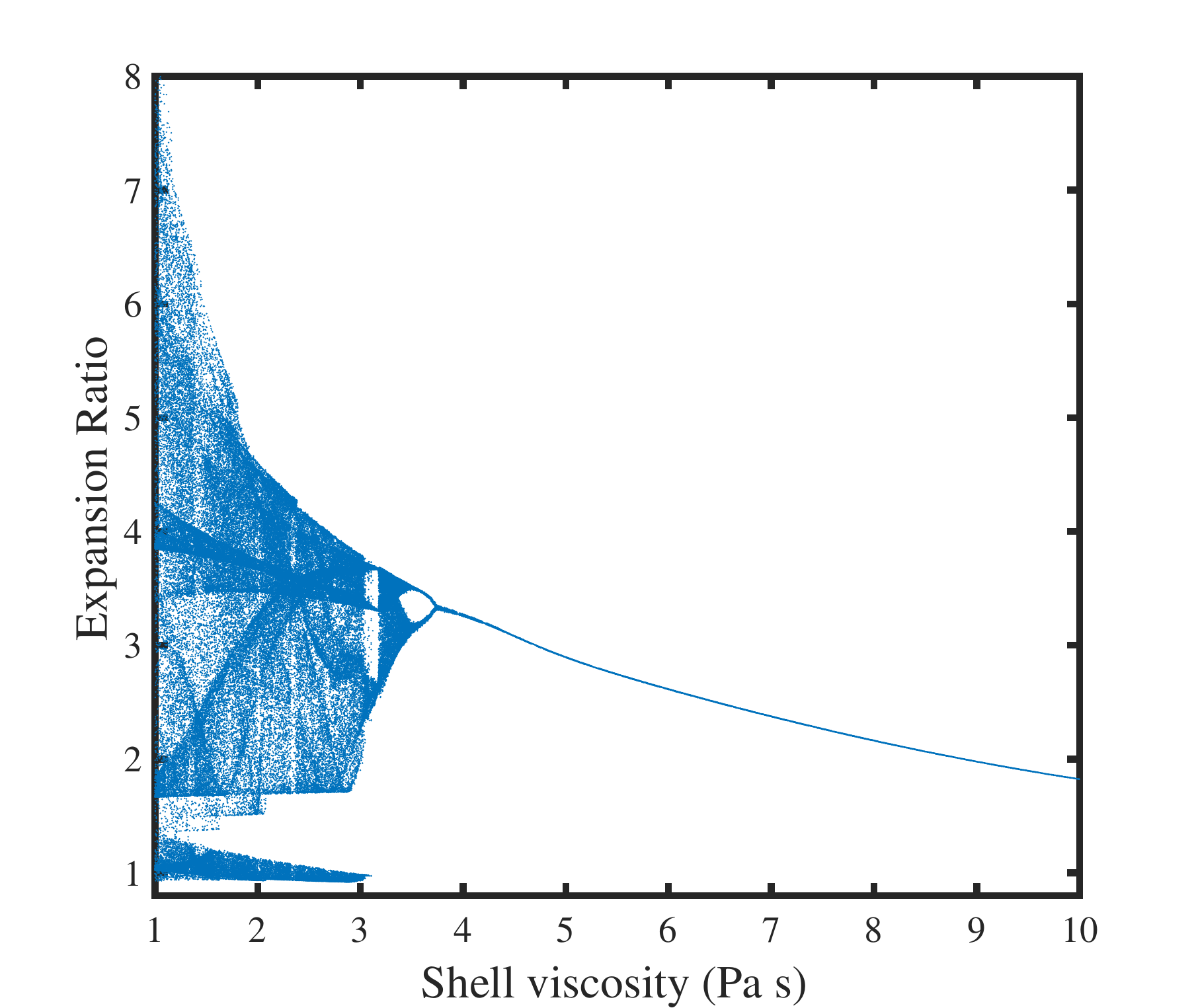}} \\
   \caption{ Bifurcation diagrams of normalized albumin-agent radius versus versus
shell viscosity when the $\nu$=1 MHz and $P_{ac}$=1.5 MPa.}
    \end{center}
\end{figure*}

\begin{figure*}
\begin{center}
 \scalebox{0.65}{\includegraphics{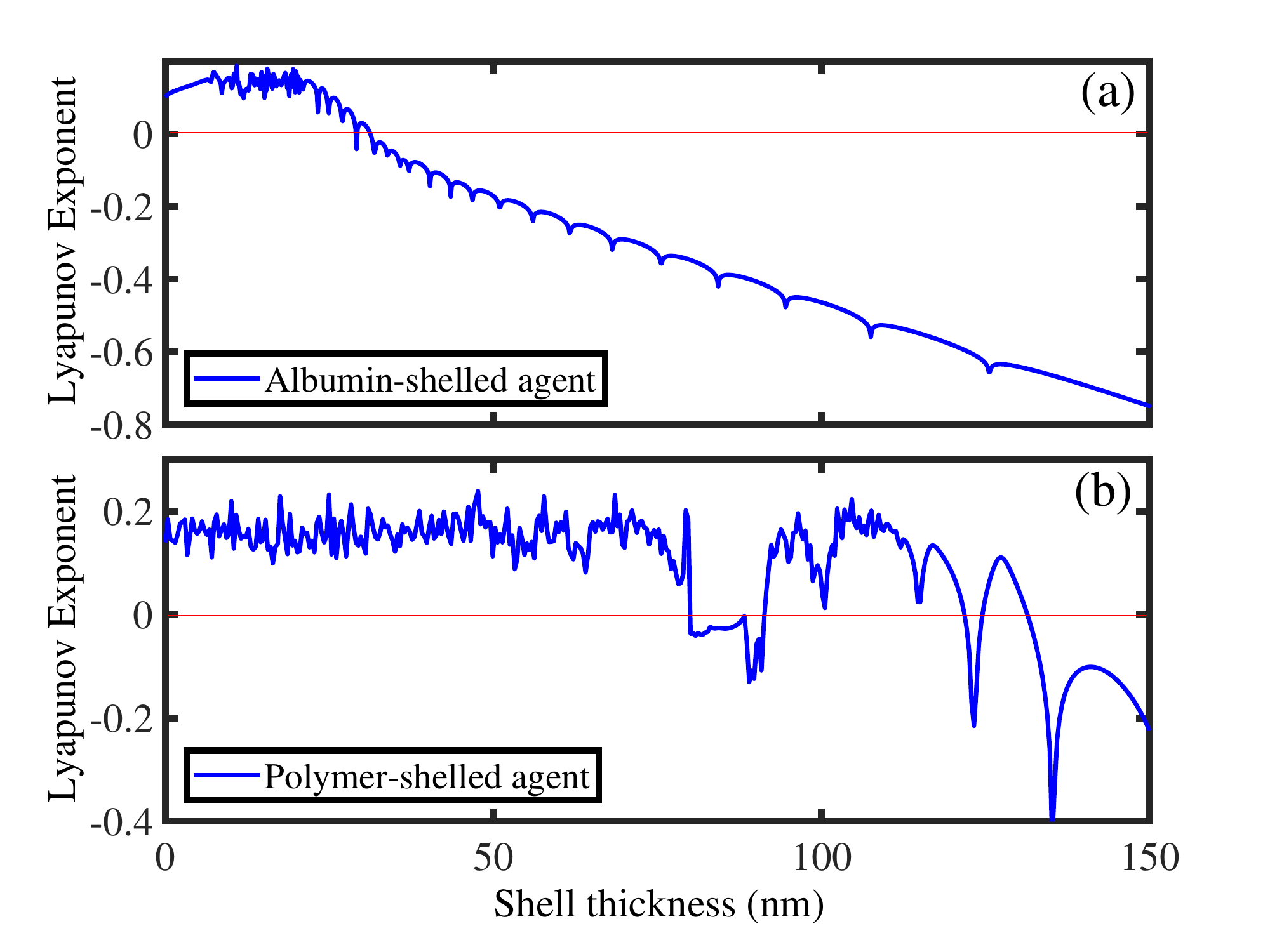}} \\
   \caption{Lyapunov spectra of normalized microbubble radius $\frac{R_{2}}{R_{20}}$ versus
shell thickness for albumin agent (a) and polymer agent (b) while the $\nu$=1 MHz and $P_{ac}$=1.5 MPa.}
    \end{center}
\end{figure*}

\begin{figure*}
\begin{center}
 \scalebox{0.65}{\includegraphics{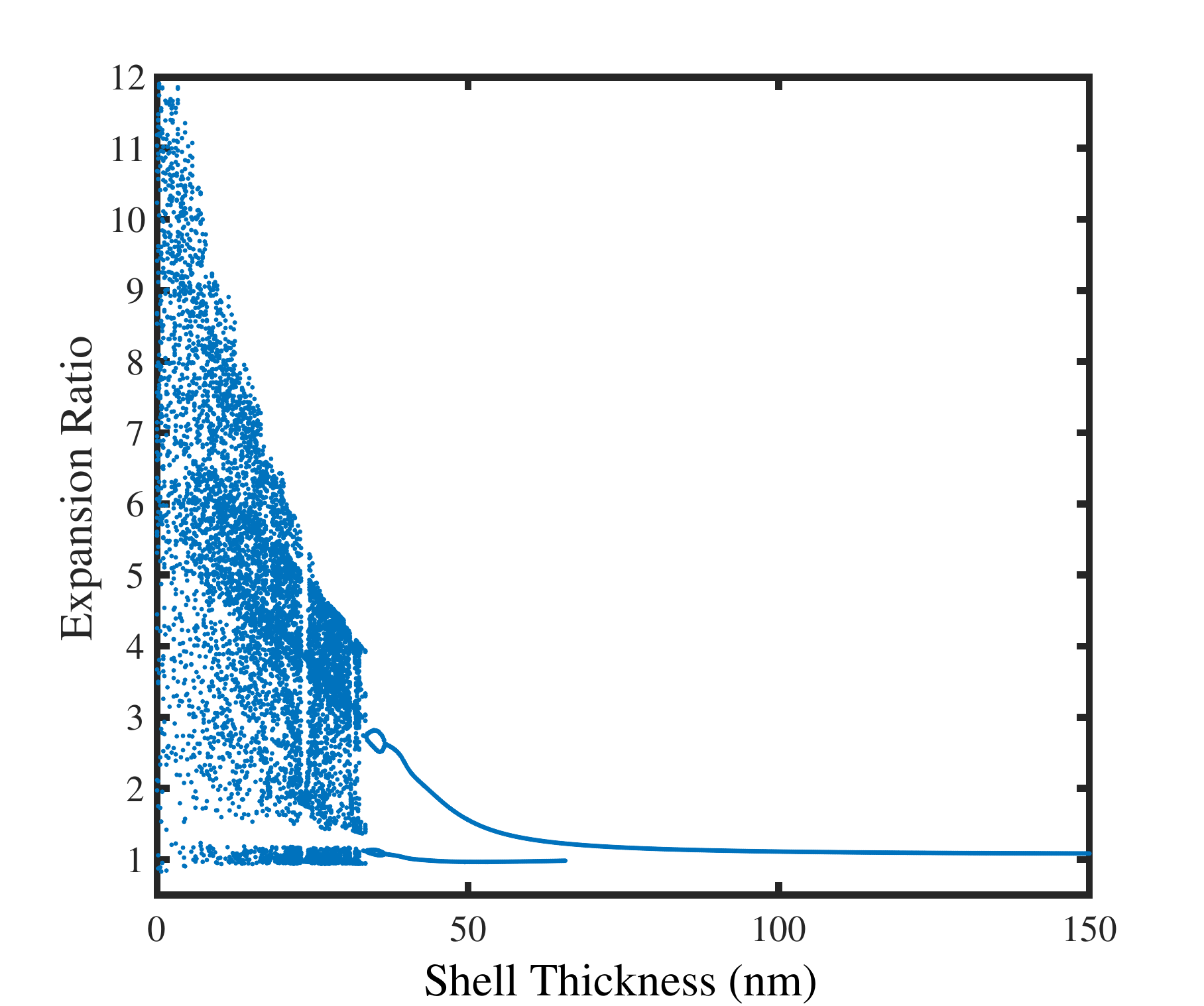}} \\
   \caption{Bifurcation diagrams of normalized albumin-agent radius versus versus
shell thickness when the $\nu$=1 MHz and $P_{ac}$=1.5 MPa.}
    \end{center}
\end{figure*}

\end{document}